\documentclass{aa}
\usepackage{graphicx}
\usepackage{txfonts}
\usepackage[]{natbib}
\newcommand{\galex}{{\it GALEX}}
\newcommand{\mi}{$\mu$m}

\newcommand{\mm}{M\,33}
\newcommand{\Ha}{H$\alpha$}

\defcitealias{2007A&A...476.1161V}{Paper I}
\defcitealias{2009A&A...493..453V}{Paper II}
\defcitealias{2009A&A...495..479C}{Paper III}
\begin{document}
   \title{On the nature of faint mid-infrared sources in M33 }

   \author{Edvige Corbelli
          \inst{1}
          \and
          Carlo Giovanardi
          \inst{1}
         \and
          Francesco Palla
          \inst{1}
          \and
          Simon Verley
          \inst{1,2}
	}

%   \offprints{}

   \institute{
Osservatorio Astrofisico di Arcetri - INAF, Largo E. Fermi 5, 50125
Firenze, Italy\\
\email{[edvige, giova, palla]@arcetri.astro.it}
\and
Dept. de F\'\i sica Te\'orica y del Cosmos, Facultad de Ciencias, 
Universidad de Granada, Spain\\
\email{simon@ugr.es}
             }

   \date{Received; accepted}

  \abstract
   {}
   {We investigate the nature of  24\mi\ sources in M33 which have weak or no 
   associated  H$\alpha$ emission. Both bright evolved stars and embedded star forming
   regions are visible as compact infrared sources in the 8 and 24\mi\ {\it Spitzer} 
   maps of M33 and contribute to the more diffuse and faint emission in these bands.
   Can we distinguish the two populations ?}
   {We carry out deep CO J=2-1 and J=1-0 line searches at the location of 18 compact 
   mid-IR  sources and 2 optically selected ones
   to unveil an ongoing star formation process throughout the disk of M33. 
   We use different assumptions to estimate cloud masses from pointed
   observations. We also analyze if the spectral energy distribution  
   and mid-IR colours  can be used to discriminate between evolved stars and 
   star forming regions. } 
   {Molecular emission  has been detected at the location of 17 sources at the
   level of 0.3~K~km~s$^{-1}$ or higher in at least one of the CO rotational lines. Even though  
   there are no giant molecular clouds beyond 4~kpc in M33, our deep observations have revealed 
   that clouds of smaller mass are very common. Estimated molecular cloud masses range between 
   10$^4$ and 10$^5$~M$_\odot$, assuming likely values of the
   CO-to-H$_2$ conversion factor and virial equilibrium. Sources which are known to be 
   evolved variable stars show weaker or undetectable CO lines.  
   Evolved stars occupy a well defined  region of the IRAC color-color diagrams. Star forming regions
   are scattered throughout a larger area even though the bulk of the distribution has different 
   IRAC colors than evolved variable stars. We estimate that about half of the 24~\mi\ sources  
   without an H$\alpha$ counterpart are genuine embedded star forming regions.
   Sources with faint but compact H$\alpha$
   emission have an incomplete Initial Mass Function (IMF) at the high-mass end and 
   are compatible with a population
   of young clusters with a stochastically sampled, universal IMF.  
   }
    {}

   \keywords{Galaxies: Individual (M\,33) --
             Galaxies: Local Group --
	     Galaxies: Star Formation --
	     ISM: Clouds --
	     ISM: Molecules --
	     Stars: AGB
            }

   \maketitle
%
%________________________________________________________________

%\tableofcontents

\section{Introduction}

Our knowledge of molecular clouds and of the processes in the interstellar medium
(hereafter ISM) that lead to the birth of stars is mostly based on Galactic studies.
Local Group galaxies are however sufficiently close to allow 
individual massive stars and molecular clouds to be detected.
M33, at a known distance, has a high star formation rate per unit area 
and a low overall extinction compared to M31, owing to the moderate gas  
content and low inclination. It is therefore an ideal laboratory  
for the investigation of the relationship of molecular clouds 
to other ISM components and evolutionary scenarios involving blue, 
low-luminosity galaxies.
Recent high-resolution optical (HST), infrared (Spitzer; hereafter, IR) and 21-cm 
observations (VLA) have traced  star 
formation and the ISM throughout the M33 disk with high accuracy.
Our investigation of the IR emission in M33 
\citep{2007A&A...476.1161V,2009A&A...493..453V,2010A&A...510A..64V, 2009A&A...495..479C}
via Spitzer high-resolution images has unveiled a variety of star formation sites 
through infrared colors and optical-to-IR ratios. 
In particular, our analysis has
shown the existence of two types of IR selected sources: sources with only diffuse
or very faint H$\alpha$ emission, and sources with a definite H$\alpha$ 
counterpart. In the former sample we can find sites  at an early stage of massive
star formation which give us  the opportunity to 
study individual embedded newly born HII regions in a galaxy different than 
our own. 

Young stellar clusters prior to the phase of gas removal,
due to photoionization or mechanical force by stellar winds, are embedded
into molecular gas and detectable only at infrared and radio wavelengths.
Previous searches in M33 have not been successful in detecting embedded 
clusters. A radio-selected sample of sources in M33 has been analyzed
by   \citet{2006ApJS..162..329B} who found optically visible counterparts with
ages between 2-10~Myr. Similar ages have been derived by \citet{2010arXiv1006.1281G}
by analyzing the spectral energy distribution of a sample of compact HII
regions. These results point out the paucity of embedded clusters which 
might be a short-lived phase of the cluster lifetime. A  dust abundance 
lower than usual or a mass spectrum of molecular clouds steeper than in our 
Galaxy might be responsible for this result. \citet{2006ApJS..162..329B}
suggest to analyze an IR selected sample, an approach that is now possible thanks
to M33 Spitzer images and 24~\mi\ source catalogue of 
\citet{2007A&A...476.1161V}.  Mid-IR sources which have no visible 
counterpart in the H$\alpha$ emission map are generally faint and their nature 
is not obvious. Candidates include 
evolved clusters, evolved stars with dusty envelopes (such as pulsating asymptotic   
giant branch, hereafter AGBs, carbon stars etc.),
embedded star forming sites, small young clusters which lack massive stars and 
hence ionizing photons. From the available catalogues we can exclude evolved
clusters  \citep[as well as Planetary Nebulae: see][]{2007A&A...476.1161V}, since
these have already removed their dusty envelope. Bright
evolved stars have also been catalogued and in this paper we discuss the likely 
presence of contamination.
The presence of molecular clouds around these sources would instead confirm 
ongoing star formation.  

A full imaging of molecular clouds complexes in M33 has been completed by the  
BIMA interferometer \citep{2003ApJS..149..343E} 
and the FCRAO-14m telescope 
\citep{2003MNRAS.342..199C,2004ApJ...602..723H} using the 
$^{12}$CO J=1-0 line.  In our Galaxy, Giant Molecular 
Clouds (hereafter GMCs) break up into smaller subunits when observed at 
high spatial resolution \citep{2003ApJ...599..258R}. 
However, in M33 both surveys  
(FCRAO and BIMA) do not find any complex above the survey completeness limit
($\sim 10^5$~M$_\odot$) beyond a galactocentric radius of 4~kpc.
On the other hand, star formation drops only at 7~kpc, 
well beyond the region where giant complexes are confined. Recent 
single dish M33 surveys \citep{2007A&A...473...91G,2010arXiv1003.3222G} have
shown that a population of low-mass molecular clouds indeed exists and becomes  
the dominant one beyond 4~kpc. Here, molecular clouds no longer aggregate into
large complexes but form predominantly in smaller mass units. This is likely 
due to the lack of spiral arms which in this flocculent spiral fade away 
around 4~kpc.
This population of molecular clouds may be more easily affected or
dispersed by the growth of HII regions and therefore their
properties and detectability might be strongly linked to the evolution of
the associated HII region. However, since the aim of the all-disk surveys was to map
large areas of the M33 disk, their sensitivity was not sufficient to
unveil the presence of molecular gas  around  most of the 
IR sources detected by Spitzer. Thus, the nature of IR sources without
associated H$\alpha$ emission needs additional efforts to be clarified.

For the aim of this paper we have restricted our sample to a few IR 
sources which are isolated, mostly located beyond 4~kpc,
in the outer regions of M33.
Our sample spans a variety of F(24$\mu$m)/F(H$\alpha$) flux ratios. The IR 
fluxes and sizes of the sources suggest that they may host small
young clusters or  OB associations, but also evolved stars.
Although there are no GMCs associated with these sources, 
the IRAM-30m telescope observations presented in this paper have given us 
the opportunity to investigate if these sources are associated to small 
molecular clouds (with masses $< 10^5$~M$_\odot$). We have carried out deep 
searches of the $^{12}$CO J=1-0 and J=2-1 rotational lines around
these sources and investigated the Spitzer IRAC 
colors of our sample and of a larger sample of AGBs and young clusters. 
There are a number of papers which investigate diagnostic IRAC color-color
diagrams to identify various type of sources 
\citep[e.g.][]{2007MNRAS.374..979C,2007A&A...476.1161V,2009ApJS..184..172G}. 
Our attempt is to 
distinguish in such color-color diagrams evolved stars from star forming regions.

The plan of the paper is the following: in
Section 2 we present the sample, the  multiwavelength data set and the CO 
observations. We discuss in Section 3 the properties of detected molecular  
clouds and in Section 4  the use of IRAC color-color diagrams
to discriminate between AGBs and HII regions. In Section 5 we use the
cluster birthline to investigate the massive stellar population in our sample. 
The conclusions are given in Section~6.

We adopt a distance to M33 of 840~kpc \citep{1991ApJ...372..455F}.

\section{Molecular gas around selected mid-IR sources}

Dust emission can be investigated through
the mid-IR and FIR data of M33 obtained with the  InfraRed Array Camera (IRAC) and 
Multiband Imaging Photometer (MIPS) on board {\it Spitzer} \citep{2004ApJS..154....1W, 
2004ApJS..154...25R,2004ApJS..154...10F}. 
The complete set of IRAC (3.6, 4.5, 5.8, and 8.0~\mi) and 
MIPS (24, and 70~\mi) images of \mm\ is described in \citet{2007A&A...476.1161V}.
The image at 70~\mi\ we use is an updated version of that of \citet{2007A&A...476.1161V}.
The total number of BCD's (Basic Calibrated Data) used in this new mosaic is
14980. The Mopex software \citep{2005ASPC..347...81M} version 18.3.1 was
used to combine the data. Data are from the PID 5 "M33 mapping and
spectroscopy" by R. Gehrz (16 AORs) and the SSC (Spitzer Science Center)
pipeline used to produce the BCDs is the version S16.1.0. The value of
the sky background removed is 9.80 (mean value of 20 median values, each
estimated in squares of 5x5 pixels, outside the galaxy). The final
measured resolution is 16" (gaussian FWHM * pixelsize = 13.3 * 1.2), as
expected. In this new
image the artificial stripes are less pronounced.
To investigate the continuum ultraviolet (UV) emission of \mm, we use 
{\it Galaxy Evolution Explorer (GALEX)} data \citep{2005ApJ...619L...1M},
in particular those distributed by \citet{2007ApJS..173..185G}. 
A description of \galex\ observations in far--UV (FUV, 1350--1750~\AA) 
and near--UV (NUV, 1750--2750~\AA) 
relative to \mm\ and of data reduction and calibration procedure 
can be found in \citet{2005ApJ...619L..67T}.

To trace ionized gas, 
we adopt the narrow-line \Ha\ image of \mm\ obtained by \citet{1998PhDT........16G}. 
The reduction process, using standard IRAF\footnote{IRAF is distributed by the 
National Optical Astronomy Observatories,
which are operated by the Association of Universities for Research
in Astronomy, Inc., under cooperative agreement with the National
Science Foundation.} procedures to subtract the continuum emission, 
is described in detail in \citet{2000ApJ...541..597H}.

\subsection{The sample}

We select a sample of 17 IR sources from the catalogue of \citet{2007A&A...476.1161V}, 
whose H$\alpha$ counterparts have flux intensities varying over two orders of magnitude.
The H$\alpha$ flux is generally modest: the brightest of these sources has a 
luminosity of 1.5$\times 10^{36}$~erg~s$^{-1}$ which corresponds to a B0 star. 
We cannot claim there is an H$\alpha$ counterpart for the faintest H$\alpha$ emission 
associated to a source but only some diffuse emission. In addition we include  
3 sources located at large galactocentric radii: one visible only in the IR (s4), 
two selected in the H$\alpha$ map (s19,s20).
In Table 1 we list the coordinates, the galactocentric distance $r$ and the emission at
various wavelengths of these sources. 
Most of them are barely resolved in the Spitzer map at 8~\mi~(2~arcsec resolution) 
and point like at 24~\mi\ (6~arcsec resolution). In the H$\alpha$ and FUV-GALEX
maps half of them are clearly visible and have a small radius of 3-5~arcsec.
If the AB magnitude in the FUV is greater than 
29.5 we cannot visually identify the source in the map.

The flux in the IRAC bands, at 24~\mi\ and in H$\alpha$ 
has been measured using a varying aperture size as described in the
source extraction algorithm of \citet{2007A&A...476.1161V}. For the UV photometry 
we use an aperture equal to the 24~\mi\ source size.
 We measure the 70~\mi\ fluxes in apertures
which are 24~arcsec in diameter (equivalent to the FWHM size of the IRAM-30~mt 
CO~J=1-0 beam). The photometric errors are small, but due to the occasional presence
of stripes in the images we can claim detection only when the flux in the aperture is 
above 8.5~mJy at 70~\mi\ ~.

\begin{table*}
\caption{Source sample and photometry in the MIR (Spitzer), UV (GALEX) and H$\alpha$.
In column 2,3 and 4 we also give the source right ascension, declination and the galactocentric
distance {\it r}.}
\label{phottab}
\begin{minipage}{\textwidth}
\begin{center}
\begin{tabular}{lcccccccccccc}
\hline \hline
     ID     & RA  &  DEC   &  $r$  & 3.6~\mi &  4.5~\mi &  5.8~\mi &  8.0~\mi &  24~\mi & 70~\mi
    &  H$\alpha \times$10$^{14}$ & FUV & NUV  \\
            & deg  & deg   &  arcmin &  mJy & mJy & mJy & mJy & mJy & mJy & erg~s$^{-1}$~cm$^{-2}$ & mag$_{AB}$ 
            &  mag$_{AB}$ \\
\hline \hline

 s1 & 23.34048 & +30.60659 &  9.20 &  0.85 &  0.99 &  1.52 &  2.85 &  3.99 &(8.2)& 0.23 & 30.4$\pm$3.9 & 29.8$\pm$1.6\\ 
 s2 & 23.35623 & +30.59784 &  8.11 &  0.77 &  1.28 &  1.54 &  1.84 &  1.35 & ... & 0.01 & .....   ...  &  .....   ...\\ 
 s3 & 23.55369 & +30.49387 & 15.17 &  1.26 &  2.40 &  3.11 &  3.56 &  3.07 & ... & 0.16 & 30.6$\pm$4.2 & 30.7$\pm$2.5\\
 s4 & 23.68463 & +30.28031 & 35.46 &  0.62 &  1.12 &  2.84 &  7.14 &  8.70 & 115 & .... & 32.4$\pm$9.6 & 31.8$\pm$4.2\\
 s5 & 23.18128 & +30.80183 & 27.06 &  0.14 &  0.06 &  0.16 &  0.57 &  3.93 & 15.5 & 0.73 & 27.8$\pm$1.1 & 28.2$\pm$0.8\\
 s6 & 23.39795 & +30.43258 & 14.29 &  0.54 &  0.35 &  1.81 &  5.65 &  7.55 & 30.7 & 7.27 & 27.6$\pm$1.1 & 27.8$\pm$0.6\\
 s7 & 23.43383 & +30.80995 & 10.92 &  0.42 &  0.22 &  1.81 &  4.59 &  4.91 & 21.6 & 3.16 & 27.9$\pm$1.2 & 28.1$\pm$0.8\\
 s8 & 23.44272 & +30.87069 & 14.59 &  0.23 &  0.18 &  0.95 &  3.18 &  6.94 & 25.5 & 5.24 & 28.7$\pm$1.7 & 28.8$\pm$1.0\\
 s9 & 23.44566 & +30.45644 & 13.18 &  0.62 &  0.29 &  0.99 &  2.75 &  6.19 & 29.8 & 21.65& 26.3$\pm$0.6 & 26.4$\pm$0.3\\
s10 & 23.49555 & +30.92542 & 16.95 &  0.19 &  0.14 &  0.71 &  2.32 & 12.55 & 18.3 & 5.44 & 28.2$\pm$1.4 & 28.4$\pm$0.9\\
s11 & 23.51050 & +30.96968 & 19.65 &  0.21 &  0.19 &  1.16 &  2.85 &  5.45 & 32.5 & 2.36 & 28.6$\pm$1.7 & 28.8$\pm$1.0\\
s12 & 23.66479 & +30.52090 & 21.20 &  0.35 &  0.16 &  2.59 &  9.37 & 21.35 & 47.7 & 1.09 & 29.0$\pm$2.0 & 29.5$\pm$1.4\\
s13 & 23.70374 & +30.62468 & 20.33 &  0.14 &  0.08 &  0.18 &  0.86 &  3.19 & 10.9 & 0.43 & 31.9$\pm$7.7 & .....   ... \\
s14 & 23.24606 & +30.69297 & 18.12 &  3.99 &  2.46 &  4.21 &  19.5 & 12.61 & 44.6 & 1.56 & 31.4$\pm$6.2 & 30.0$\pm$1.8\\
s15 & 23.55395 & +30.45810 & 17.34 &  0.26 &  0.19 &  0.38 &  4.85 & 14.30 & 21.1 & 0.82 & 30.6$\pm$4.3 & 30.4$\pm$2.2\\
s16 & 23.56527 & +30.46270 & 17.69 &  1.66 &  1.23 &  1.30 &  4.11 & 11.23 & 20.9 & 0.45 & 29.7$\pm$2.7 & 29.5$\pm$1.4\\
s17 & 23.61698 & +30.93344 & 18.30 &  0.20 &  0.25 &  0.52 &  1.85 &  7.33 & 12.5 & 0.28 & 30.6$\pm$4.3 & 31.5$\pm$3.5\\
s18 & 23.69661 & +30.63216 & 19.49 &  0.30 &  0.17 &  1.23 &  5.31 &  5.09 & 31.7 & 4.57 & 27.9$\pm$1.2 & 28.1$\pm$0.8\\
s19 & 23.67558 & +30.40020 & 28.08 &  0.08 &  0.06 &  0.30 &  0.95 &   0.9 & (8.1)& 1.8  & 29.6$\pm$2.6 & 29.4$\pm$1.4\\
s20 & 23.74792 & +30.61717 & 23.99 &  0.54 &  0.71 &  0.64 &  0.95 &   ... & ...  & 4.0  & 27.8$\pm$1.2 & 27.7$\pm$0.6\\

\hline \hline
\end{tabular}
\end{center}
\end{minipage}
\end{table*}

\subsection {Contamination by AGB stars}

Since 24~\mi\ sources can be star forming regions or evolved stars 
\citep[see ][ ]{2009A&A...493..453V},
we now check whether some of the sources in our sample have been catalogued as evolved variable 
stars. We cross checked our source list with long period
variable stars catalogued by \citet{2007ApJ...664..850M} from MIR observations and with the
variable point source catalogue of \citet{2006MNRAS.371.1405H} at optical wavelengths. We
set the searching radius equal to the size of the 24~\mi\ emission.
We found that the first three infrared selected sources in our list (s1,s2,s3) are variables 
according to the classification
scheme of \citet{2007ApJ...664..850M}. The angular separation between the variable star and 
the center of the 24~\mi\ emission is less than 1.3~arcsec, i.e. 5~pc. We also found that the 
last source in our sample 
(s20, optically selected) is in the variable point source list of \citet{2006MNRAS.371.1405H}.
Given the conspicuous H$\alpha$ emission associated with this region, we believe that
the source s20 is an example of an evolved star close to a
young cluster (at a distance of about 2~pc).
Given its spectral characteristics, which will be shown later in the paper, s4 might also  
be an evolved star: a galactic AGB or an AGB in M33 at a large galactocentric distance, 
beyond the area surveyed by \citet{2007ApJ...664..850M}. 
We keep the evolved stars in our sample to check if there is any detectable CO
emission from the surrounding region.

\subsection {The IRAM-30mt observations}

Some CO J=1-0 emission has already been detected with a 45~arcsec wide
beam (FCRAO) around a few sources of our sample \citep{2004ApJ...602..723H}. 
Given the large beam size, 
we do not know if the detected gas is associated to or in 
the proximity of the sources. Therefore, we have searched for CO emission using the smaller IRAM-30~mt beam 
from all the sources in Table 1. The CO J=1-0 and J=2-1 lines have been observed 
during August 2007 with 
a FWHM beam of 24~arcsec at 115~GHz and of 12~arcsec at 230~GHz.
At 24~\mi\  all the sources are smaller in size than the telescope beam at 230~GHz.

We have observed the sources in position switching mode,
using the receiver combination A100/B100 and A230/B230
and the VESPA backend system with  240 MHz bandwidth (320 kHz resolution).
One source was centered on the ON position, another source in the OFF
position. The OFF source is chosen in
a region with different line of sight velocity than the ON position
(as seen through 21-cm maps).
The spectra have been smoothed to 1~km~s$^{-1}$ and the data from both 
receivers has been averaged. 

In Table 2 we summarize the CO data: integrated emission
I (in units of main beam temperature  K~km~s$^{-1}$), mean velocity V, 
line width W ( full width at half maximum; hereafter, FWHM) and peak intensity P,
are estimated using gaussian fits to the lines. The rms refers to a spectral resolution of
2.2 and 1.1~km~s$^{-1}$ for the CO J=1-0 and J=2-1, respectively.  The line widths have been
measured using correlator spectra after correcting for hanning.

\begin{table*}
\caption{ Gaussian fits to CO lines. In column 1 and 2 we show the integrated 
intensity of the J=1-0 and J=2-1 lines respectively, in column 3 and 4 the mean velocity of 
the two lines, in column 5 and 6 the relative line widths, in column 7 and 8 the peak intensities, 
and in the last two columns the rms for the CO J=1-0 and J=2-1 respectively.}
\label{lines}
\begin{minipage}{\textwidth}
\begin{center}
\begin{tabular}{lcccccccccc}
\hline \hline
ID     & I$_{1-0}$  & I$_{2-1}$ &  V$_{1-0}$  & V$_{2-1}$ &  W$_{1-0}$  & W$_{2-1}$ & P$_{1-0}$  & P$_{2-1}$ 
      & rms$_{1-0}$ & rms$_{2-1}$ \\
      & K~km~s$^{-1}$ & K~km~s$^{-1}$ &  km~s$^{-1}$ &  km~s$^{-1}$ & km~s$^{-1}$ & km~s$^{-1}$ & K & K 
      & K & K \\
\hline \hline

s1  & 0.284$\pm$0.028 & 0.665$\pm$0.037 & -156.87$\pm$0.20 & -156.91$\pm$0.10 & 2.90$\pm$0.48 & 3.19$\pm$0.31
    & 0.063 & 0.165 & 0.008 & 0.019 \\
s2  & $<$0.150 & $<$0.200 & & & & & & & 0.009 & 0.018 \\ 
s3  & 0.247$\pm$0.025 & 0.467$\pm$0.047 & -155.82$\pm$0.35 & -155.31$\pm$0.64 & 5.75$\pm$1.69 & 11.67$\pm$2.35  
    & 0.039 & 0.036 & 0.008 & 0.015 \\
s4  & $<$0.100 & $<$0.150 & & & & & & & 0.007 & 0.014 \\ 
s5  & 0.665$\pm$0.038 & 0.746$\pm$0.107 & -179.38$\pm$0.16 & -178.93$\pm$0.29 & 4.93$\pm$ 0.65 & 3.79$\pm$0.59  
    & 0.109 & 0.181 & 0.011 & 0.059 \\
s6  & 0.303$\pm$0.022 & 0.639$\pm$0.038 & -108.04$\pm$0.10 & -107.89$\pm$0.08 & 1.61$\pm$0.29  & 1.89$\pm$0.29 
    & 0.106 & 0.222 & 0.010 & 0.024 \\
s7  & 1.011$\pm$0.025 & 1.338$\pm$0.033 & -240.99$\pm$0.07 & -240.94$\pm$0.06 & 4.99$\pm$0.27 & 4.97$\pm$0.19  
    & 0.162 & 0.244 & 0.007 & 0.015 \\  
s8  & 0.483$\pm$0.028 & 0.981$\pm$0.035 &  -243.49$\pm$0.13 & -243.39$\pm$0.07 &  4.03$\pm$0.50 & 3.33$\pm$0.21 
    & 0.082 & 0.235 & 0.007 & 0.020 \\   
s9  & 1.206$\pm$0.028 & 1.921$\pm$0.035 & -124.46$\pm$0.05 & -124.36$\pm$0.03 & 3.93$\pm$0.19 & 3.37$\pm$0.10  
    & 0.235 & 0.494 & 0.009 & 0.017 \\             
s10 & 0.646$\pm$0.023 & 1.762$\pm$0.034 & -253.66$\pm$0.08 & -253.87$\pm$0.04 &  3.61$\pm$0.22 &  3.87$\pm$0.13 
    & 0.134 & 0.386 & 0.008 & 0.017 \\    
s11 & 1.114$\pm$0.027 & 0.704$\pm$0.039 & -261.80$\pm$0.10 & -261.87$\pm$0.14 & 7.64$\pm$0.65 & 4.35$\pm$0.63
    & 0.126 & 0.128 & 0.007 & 0.017 \\
s12 & 1.308$\pm$0.051 & 1.516$\pm$0.042 &  -151.74$\pm$0.12 & -151.40$\pm$0.06 & 5.54$\pm$0.29 & 5.30$\pm$0.30  
    & 0.196 & 0.290 & 0.007 & 0.018 \\   
s13 & 0.809$\pm$0.024 & 1.001$\pm$0.039 & -180.78$\pm$0.07 & -180.06$\pm$0.07 & 3.89$\pm$0.21 & 3.48$\pm$0.23
    & 0.163 & 0.249 & 0.008 & 0.021 \\                       
s14 & 1.326$\pm$0.026 & 1.191$\pm$0.035 & -173.56$\pm$0.05 & -173.50$\pm$0.06 & 4.92$\pm$0.25 & 3.90$\pm$0.24  
    & 0.211 & 0.267 & 0.007 & 0.017 \\ 
s15 & 0.374$\pm$0.021 & 0.406$\pm$0.027 & -137.83$\pm$0.11 & -137.45$\pm$0.09 & 3.11$\pm$0.66 & 2.19$\pm$0.40  
    & 0.088 & 0.133 & 0.008 & 0.017 \\   
    & 0.150$\pm$0.020 & 0.295$\pm$0.026 & -130.72$\pm$0.25 & -131.17$\pm$0.12 & 2.33$\pm$1.45 & 2.36$\pm$0.49 
    & 0.039 & 0.108 & 0.008 & 0.017 \\
s16 & $<$0.100 & $<$0.150 & & & & & & & 0.007 & 0.017\\ 
s17 & 1.10$\pm$0.029 & 0.838$\pm$0.058 & -261.31$\pm$0.09 & -262.45$\pm$0.16 & 6.49$\pm$0.40 & 4.61$\pm$0.44             
    & 0.141 & 0.171 & 0.008 & 0.028 \\      
s18 & 0.975$\pm$0.030 & 1.259$\pm$0.036 & -183.14$\pm$0.08 & -182.97$\pm$0.06 & 5.73$\pm$0.43 & 4.65$\pm$0.23
    & 0.148 & 0.252 & 0.008 & 0.016 \\   
s19 & 0.666$\pm$0.067 & 0.654$\pm$0.065 & -152.91$\pm$0.49 & -154.16$\pm$0.35 & 4.80$\pm$1.08 & 6.32$\pm$0.91 
    & 0.059 & 0.093 & 0.014 & 0.028  \\ 
s20 & 0.344$\pm$0.050 & 0.178$\pm$0.044 & -187.52$\pm$0.57 & -190.32$\pm$0.32 & 4.03$\pm$1.15 & 2.03$\pm$0.86
    & 0.046 & 0.064 & 0.013 & 0.029  \\  

\hline \hline
\end{tabular}
\end{center}
\end{minipage}
\end{table*}

We shall use as CO J=1-0 line intensity the average value of that derived by fitting a gaussian  to
the detected emission and that obtained by summing the flux in each channel inside the signal window 
(determined
individually for each spectrum). The result is given in Table 3.
As uncertainties on the CO line intensities, we consider the largest value between
the following ones: uncertainty derived from the gaussian fit, the rms of the spectra
integrated over the signal window, the dispersion between the
intensity derived from gaussian fit, and the intensity derived by the integral 
inside the window signal.
The gain is 6.3 and 8.7 Jy~K$^{-1}$ at 110 and 235~GHz, respectively. Using these values,
we derive the CO fluxes of the two rotational levels and their ratios.
In Table 3 we also quote  the H$_2$ column densities derived from FCRAO (beam=45~arcsec)
integrated CO J=1-0 fluxes, using a CO-H$_2$ conversion factor 
X$_{CO}$=2.8~10$^{20}$~cm$^{-2}$~K$^{-1}$. If there are
no entries in the column corresponding to the FCRAO column density, it means that the source was off the
region mapped by FCRAO.

The bolometric luminosity for star forming regions in Table 3 is computed as
the sum of the FUV and NUV luminosities
uncorrected for absorption, added to the total infrared luminosity (hereafter, TIR luminosity):

\begin{equation}
L_{bol}=L_{FUV}+L_{NUV}+L_{TIR}+24 L_{H\alpha}
\end{equation}

\noindent
where the H$\alpha$ luminosity term accounts for the  
ionizing radiation \citep{1999ApJS..123....3L} 
and L$_{TIR}$ for the UV radiation absorbed by grains and re-emitted in the
IR. We have not considered the continuum radiation longward of 2800~\AA
which becomes important when young clusters have luminosities lower than 10$^{38}$~erg~s$^{-1}$.  
The estimated TIR luminosity, $L_{TIR}$,  has been computed 
following \citet{2009A&A...493..453V}: 

\begin{equation}
{\hbox{log}} L_{TIR}=log L_{24} + 1.08 + 0.51\times {\hbox{log}} {L_{\nu,8} \over L_{\nu,24}}
\end{equation}

\noindent
and the luminosity function from the 8 and 24~\mi\ fluxes as  $\nu L_\nu$. The above expression 
is correct for HII
regions whose IR emission peaks longward of 24~\mi\. Evolved stars s1,s2,s3 have no
emisson in the FIR and the above formula overestimates the TIR. For these sources we 
compute the luminosity according to \citet{2007MNRAS.376..313G} who have shown that the dust
emission from evolved stars closely follows that of a black body at temperatures of 400-600~K. 
In particular, using the 8 and 24~\mi\ fluxes given in Table~1, we estimate the luminosity 
in solar luminosity units as:

\begin{equation}
L_{bol}(AGB)=130\ {F_8\over mJy} \big({F_8\over F_{24}}\big)^{1.76} 
exp(4.4\ (F_{24}/F_8)^{0.44})
\end{equation}

\noindent
For s4 we give the luminosities and stellar masses using the formulae for HII regions
but put the values in brackets since it might be an evolved star.
Extinction corrections for H$\alpha$ fluxes in HII regions are 0.83 A$_V$ where A$_V$,
the visual extinction, is given as  

\begin{equation}
A_V=3\times 0.57\times {\hbox{log}}\bigl({L_{TIR}\over L_{NUV} + L_{FUV}}+1\bigr)
\end{equation}

In Table 3 we list extinction values and cluster masses. 
Following \citet{2009A&A...495..479C}, we set 0.1~M$_\odot$ as the lower limit of the IMF  
and use a Salpeter slope of 2.3 down to 0.5~M$_\odot$, and 1.3 between 0.5
and 0.1~M$_\odot$. 
When bolometric luminosities are small, as in the case of these clusters, the upper 
end of the IMF is not fully populated and one has to take into account stochastic
effects. In the stochastic regime, uncertainties are large and
a simple scaling law between luminosity and cluster mass
does not apply (and it would underestimate the cluster mass). 
For L$_{bol}<10^{40}$~erg~s$^{-1}$ the possible range of cluster masses for a given
L$_{bol}$ is up to one order of magnitude \citep{2010arXiv1011.1097C}. In Table 3 we show the 
median cluster mass for the given bolometric luminosity,  the expected
median H$\alpha$ luminosity, L$^{exp}_{H\alpha}$ (corresponding to the observed L$_{bol}$) 
and the relative uncertainties. The observed values corrected for extinction, 
L$^{obs}_{H\alpha}$,  are also given. Finally, we comment  
when there is not a clear peak in the images at the location of the 24~\mi\ source.

\begin{table*}
\caption{Brightness of CO clouds and associated stellar clusters.
In columns 2 and 3 we give the value of the CO J=1-0 and 2-1 line intensity averaged between 
that derived from gaussian fits and that obtained by summing the flux in each channel. 
The ratio between the two CO line intensities is shown in column 
4. In column 5 we give the H$_2$ column densities derived from FCRAO observations and
in column 6 and 7 the H$\alpha$ luminosity (corrected for extinction) and the bolometric
luminosity of the sources. In the last 3 columns we 
show the estimated mass, expected H$\alpha$ luminosity (see text for details),
and visual extinction of the stellar clusters.}

\label{masstab}
\begin{minipage}{\textwidth}
\begin{center}
\begin{tabular}{lccccccccccc}
\hline \hline

     ID     & $<$I$_{1-0}>$ & $<$I$_{2-1}>$  & ${<I_{2-1}>\over <I_{1-0}>}$ & log N$_{H_2 FCRAO}^{1-0}$
            & log L$_{H\alpha}^{obs}$ & log L$_{bol}^{obs}$ & M$_{*}$ & log L$^{exp}_{H\alpha}$ & A$_V$ \\
            & K~km~s$^{-1}$ & K~km~s$^{-1}$ &   & cm$^{-2}$
            & erg~s$^{-1}$ & erg~s$^{-1}$ & M$_\odot$ & erg~s$^{-1}$ & mag  & Notes\\
            
\hline \hline

     s1 & 0.300$\pm$0.028 &  0.658$\pm$0.037 & 2.19 &  $<20.4$ & 36.11 & 38.13$^{a}$ & ... & ... & ...  & Visible only at 8-24-70~\mi\\
     s2 & $<$0.150        & $<$0.200         & ...  &  $<20.4$ & $<$33.97& 37.88$^{a}$ & ... & ... & ...  & Visible only at 8-24~\mi\\
     s3 & 0.265$\pm$0.025 &  0.306$\pm$0.228 & 1.15 &  $<20.4$ & 35.99 & 38.17$^{a}$ & ... & ... & ... & Visible only at 8-24~\mi\\
     s4 & $<$0.100        & $<$0.150         & ...  &   ...    & 34.56 & (39.00)$^{a}$ & (490) & (37.1$^{+0.7}_{-1.5}$) & 4.2  & Visible only at 8-24-70~\mi\\
     s5 & 0.655$\pm$0.038 &  0.718$\pm$0.107 & 1.10 & ...      & 35.88 & 38.64 & 258 & 36.4$^{+1.1}_{-2.9}$ & 0.4 & Weak H$\alpha$,non isol.\\
     s6 & 0.290$\pm$0.022 &  0.571$\pm$0.097 & 1.97 & $<20.4$  & 37.11 & 39.15 & 582 & 37.2$^{+0.7}_{-1.1}$ & 1.0 & \\
     s7 & 1.060$\pm$0.069 &  1.363$\pm$0.035 & 1.29 & 21.02$^{+0.09}_{-0.12}$& 36.74 & 38.98 & 431 & 37.0$^{+0.8}_{-2.0}$ & 0.9 & \\
     s8 & 0.565$\pm$0.115 &  0.947$\pm$0.047 & 1.68 & 21.18$^{+0.10}_{-0.14}$& 37.13 & 39.00 & 490 & 37.1$^{+0.7}_{-1.5}$ & 1.4 & \\
     s9 & 1.229$\pm$0.033 &  1.861$\pm$0.085 & 1.51 & $<20.4$  & 37.32 & 39.30 & 794 & 37.4$^{+0.6}_{-0.8}$ & 0.3 & \\
     s10& 0.642$\pm$0.023 &  1.765$\pm$0.034 & 2.75 & 20.90$^{+0.11}_{-0.15}$& 37.07 & 39.04 & 504 & 37.1$^{+0.7}_{-1.5}$ & 1.2 & \\
     s11& 1.087$\pm$0.038 &  0.659$\pm$0.064 & 0.61 & ...                    & 36.73& 38.86 & 370 & 36.9$^{+0.8}_{-2.4}$ & 1.2 & \\
     s12& 1.282$\pm$0.051 &  1.681$\pm$0.283 & 1.31 & 20.65$^{+0.16}_{-0.26}$& 36.83 & 39.30 & 794 & 37.4$^{+0.6}_{-0.8}$ & 2.3 & No FUV, tail\\
     s13& 0.888$\pm$0.112 &  1.000$\pm$0.039 & 1.13 & 21.23$^{+0.05}_{-0.07}$& 36.58 & 38.47 & 215 & 36.0$^{+1.3}_{-2.8}$ & 2.7 & No FUV\\
     s14& 1.422$\pm$0.135 &  1.263$\pm$0.102 & 0.89 & 20.78$^{+0.15}_{-0.22}$& 37.69 & 39.49 & 948 & 37.6$^{+0.4}_{-0.7}$ & 4.1 & No FUV\\
     s15& 0.372$\pm$0.020 &  0.352$\pm$0.016 & 0.95 & $<20.4$  	             & 36.99 & 39.11 & 566 & 37.2$^{+0.7}_{-1.2}$ & 3.0 & No FUV, weak H$\alpha$\\
     s15& 0.159$\pm$0.020 &  0.315$\pm$0.028 & 1.98 & $<20.4$                & 36.99 & 39.11 & 566 & 37.2$^{+0.7}_{-1.2}$ & 3.0 & No FUV, weak H$\alpha$\\
     s16& $<$0.100        & $<$0.150         &  ... & $<20.4$                & 36.41 & 38.98 & 431 & 37.0$^{+0.8}_{-2.0}$ & 2.3 & Weak H$\alpha$\\
     s17& 1.159$\pm$0.082 &  0.852$\pm$0.058 & 0.74 & 21.13$^{+0.11}_{-0.15}$& 36.29 & 38.72 & 313 & 36.5$^{+1.1}_{-2.7}$ & 2.5 & Visible only at 8-24-70~\mi\\
     s18& 1.012$\pm$0.052 &  1.269$\pm$0.036 & 1.25 & 20.60$^{+0.21}_{-0.39}$& 36.93 & 39.03 & 500 & 37.1$^{+0.7}_{-1.5}$ & 1.0 & \\
     s19& 0.651$\pm$0.067 &  0.637$\pm$0.065 & 0.98 & ...                    & 36.48 & 38.38 & 206 & 35.8$^{+1.5}_{-2.7}$ & 0.9 & IR weak\\
     s20& 0.353$\pm$0.050 &  0.172$\pm$0.044 & 0.49 & ...                    & 36.48 & 38.54 & 241 & 36.3$^{+1.2}_{-2.8}$ & 0.04 & IR weak\\

\hline \hline
\end{tabular}
\end{center}
{$^{a}$ Luminosity as from equation (3), since source is an AGB star. $^{b}$ Source of unknown
nature, luminosity computed according to equation (1).  $^{c}$ HII region next to an AGB star,
luminosity computed according to equation (1)}.
\end{minipage}
\end{table*}

In the next Section and in the rest of this Section we discuss possible correlations
between cloud properties. We will quote in parenthesis the Pearson linear correlation 
coefficient $r_P$ for data relative to star forming regions, excluding AGBs.

The ratio of line intensities $I_{2-1}/I_{1-0}$ in Figure~1
correlates with the ratio of line-widths ($r_P=0.74$ for  
log $W_{2-1}/W_{1-0}$ -- log $I_{2-1}/I_{1-0}$ ). If we exclude s19, the source
with the largest $W_{2-1}/W_{1-0}$, the Pearson linear correlation coefficient is
0.91 and the slope is 1.9$\pm 0.2$. The observed line ratio shows a marginal decrease as the 
galactocentric radius increases ($r_P=-0.46$) and it does not correlate with the cluster 
bolometric luminosity.
In Figure~1 we can see also that the emission at 70~\mi\ increases with the cluster mass, as
expected, since for most of our clusters the infrared luminosity drives the value of
the bolometric luminosity (even though we have not used the 70~\mi\ emission to compute the
TIR). The scatter is however large ($r_P=0.46$).
The brightness of the CO lines does not correlate with the median cluster mass
($r_P=0.38$), even though Figure~1 shows that
more massive clusters are associated with the brightest CO J=2-1 lines.

\begin{figure}
\includegraphics[width=\columnwidth]{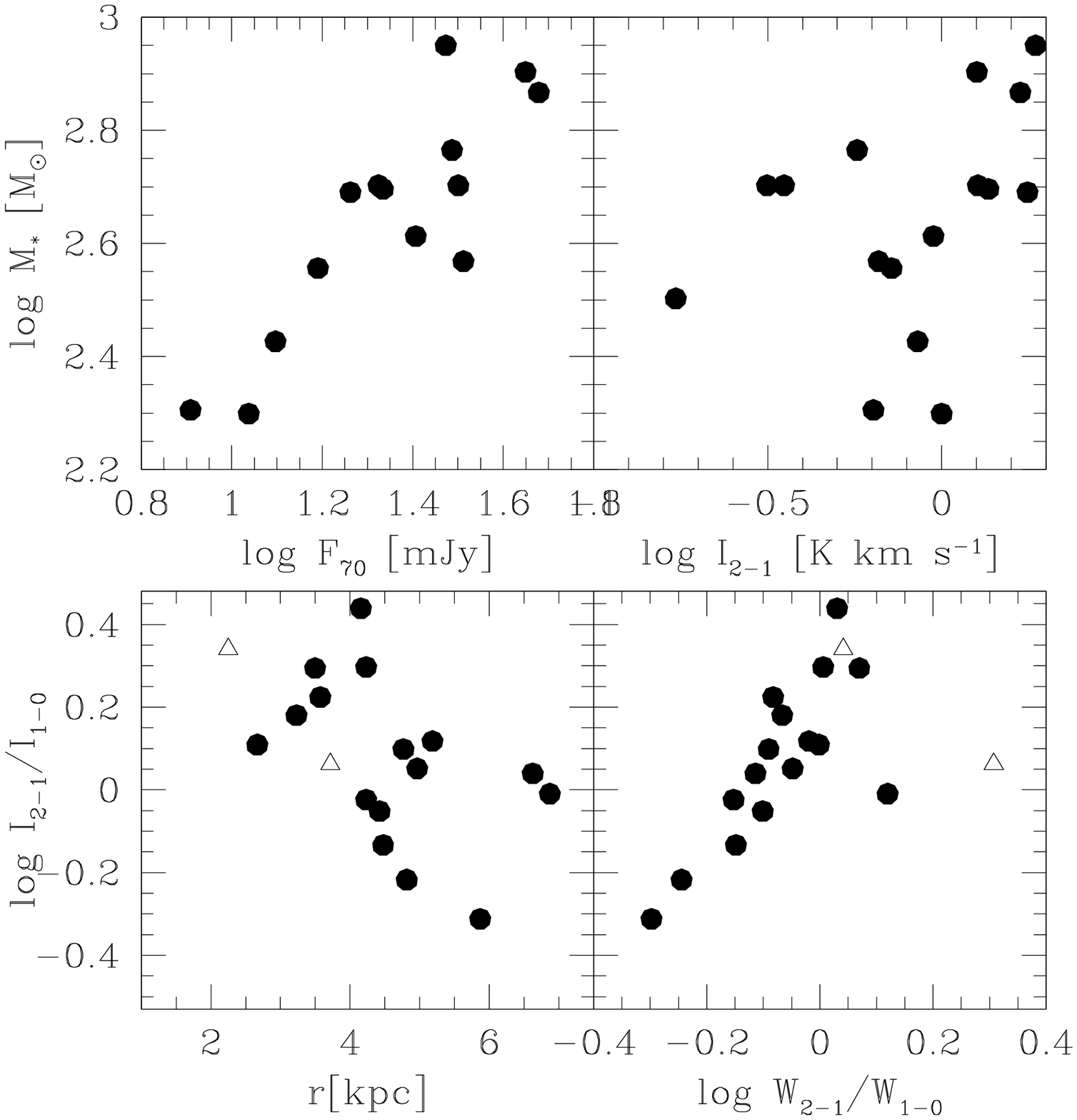}
\caption{The correlation of F$_{70}$ and the cluster mass is shown in the
upper left panel. The upper right panel shows that clusters of larger mass are associated with
brighter CO J=2-1 lines.  
The bottom panels display the marginal correlation between the line intensity 
ratio $I_{2-1}/I_{1-0}$  and the galactocentric radius and the good correlation between 
$I_{2-1}/I_{1-0}$
and the line width ratio $W_{2-1}/W_{1-0}$. Triangles are for variable stars (s1,s3) and are
displayed only in the lower panels. }
\label{fig:f1}
\end{figure}

\section {Three methods for computing molecular masses and sizes}

We compute the properties of the molecular gas following 3 different methods. 
In practice, we have three unknowns: the cloud size, the
conversion factor $X_{CO}$ between the CO luminosity and the H$_2$ surface density,
and $R$ the intrinsic line ratio. The latter quantity is the
ratio between the J=2-1 and J=1-0 lines we would measure by mapping the cloud with 
a telescope beam smaller than the cloud
size. The expected line ratio is the line ratio we expect to observe knowing the
telescope beam size at the two line frequencies, the intrinsic line ratio and the source 
extent. The observed line ratio is the measured value of I$_{2-1}$/I$_{1-0}$.
If we consider the cloud virialized and centered at the source location, we have 2 equations 
for these 3 variables: one equation is given by equating
the observed to the expected line ratio and the other one by equating the virial mass to the 
mass inferred from the CO line brightness. Clearly, even in this case some assumption is needed
to derive the cloud properties such as mass, density, size etc.. We now describe 3 different
methods to derive cloud parameters and the underlying assumptions.

\subsection  {Uniform intrinsic line ratio}

We assume a fixed intrinsic line ratio $R$ between the CO J=2-1 and 1-0 lines for all sources
and compute the expected line ratio
under the hypothesis that the source is isolated and centered 
in the beam. We infer the source diameter $D$ by equating the expected
ratio to the observed one:

\begin{table*}
\caption{Cloud parameters for an intrinsic line ratio $R=0.5$. The cloud diameter  is given
in column 2 and the virial masses corresponding to the J=1-0 and J=2-1 CO lines are given in
column 3 and 4. The H$_2$ volume and column density are shown in column 5 and 6
respectively and the CO-to-H$_2$ conversion factor in the last column (average value between 
those relative to the two CO lines).}
\label{par3}
\begin{minipage}{\textwidth}
\begin{center}
\begin{tabular}{lccccccc}
\hline \hline
     ID  & D &   M$_{vir}^{1-0}$ & M$_{vir}^{2-1}$ & n$_{H_2}^{1-0}$  & N$_{H_2}$ & X$_{CO}$ \\
         & pc  & 10$^4$ M$_\odot$ & 10$^4$M$_\odot$ & cm$^{-3}$ & 10$^{20}$& 10$^{20}$~cm$^{-2}$~K$^{-1}$\\
\hline \hline
    s1 &  ..& ..  & .. &  .. & .. & ..\\
    s2 &  ..& ..  & .. &  .. & .. & ..\\
    s3 & 77 &  26.&107.&   15& 36 & 58\\
    s4 &  ..& ..  & .. &  .. & .. & ..\\
    s5 & 81 &  20.& 12.&   10& 25 & 5.6\\
    s6 & 12 & 0.31&0.42&   54& 20 & 0.7\\
    s7 & 69 &  17.& 17.&   15& 32 & 4.0\\
    s8 & 41 &  6.7& 4.5&   28& 35 & 2.4\\
    s9 & 53 &  8.2& 6.0&   16& 26 & 1.6\\
   s10 &  ..& ..  & .. &  .. & .. & ..\\
   s11 & 155&  89.& 29.&    7& 33 & 7.3\\
   s12 & 57 &  18.& 13.&   27& 47 & 3.2\\
   s13 & 81 &  12.& 9.7&    7& 18 & 2.9\\
   s14 & 102&  25.&15.6&    7& 22 & 3.0\\
   S15 & 98 &  9.3& 4.6&    4& 12 & 4.0\\
   s15 &  8 & 0.52&0.54&  168& 41 & 1.0\\
   s16 &  ..& ..  & .. &  .. & .. & ..\\
   s17 & 114&  48.& 24.&    9& 32 & 6.0\\
   s18 & 73 &  23.&15.4&   19& 43 & 2.8\\
   s19 & 94 &  22.& 37.&    8& 23 & 9.6\\
   s20 &  ..& ..  & .. &  .. & .. & ..\\

\hline \hline
\end{tabular}
\end{center}
\end{minipage}
\end{table*}

\begin{equation}
R={I_{2-1} (1-exp(-D^2/(1.2\Theta_{1-0})^2))\over I_{1-0} (1-exp(-D^2/(1.2\Theta_{2-1})^2))}
\end{equation}

\noindent
where $\Theta$ is the half-width half-maximum of the beam. Knowing the size
we can infer the virial mass and by equating the virial mass 
to the mass derived using the CO line luminosity (which
is the same for the 2-1 and 1-0 line once we correct for beam dilution), we infer $X_{CO}$.
Other cloud properties, such as the H$_2$ volume or column density, then follow.
The cloud virial mass for a density profile which varies inversely to the cloud
radius can be written as \citep{1990ApJ...363..435W}:

\begin{equation}
{M_{vir}\over M_\odot} = 99 {D\over pc} ({W\over km~s^{-1}})^2 = 406 {D\over arcsec} ({W\over km~s^{-1}})^2
\end{equation}

\noindent
where W is the FWHM of the CO line and D is the diameter or cloud size. 
The average volume density is 

\begin{equation}
{n_{H_2}\over cm^{-3}} = {406\ D_{arcsec}\ W_{km~s^{-1}}^2\ 2\times 10^{33}\over D_{arcsec}^3 
1.05\times 10^{57}\ 1.33\times m_{H_2}} = 174 {W_{km~s^{-1}}^2\over D_{arcsec}^2 }
\end{equation}

\noindent
where the factor 1.33 accounts for helium.
The average intrinsic line ratio in molecular clouds of our Galaxy varies between 
0.5 and 0.8 \citep{1998ApJ...493..730O}, being higher at low galactic latitudes. In external 
galaxies intrinsic ratios  are as low as 0.3 or as high as 2  
\citep[e.g.][ ]{2001ApJ...551..794S,2007AJ....134.1827C} and can be explained in terms
of variations of the cloud physical conditions \citep{1994ApJ...425..641S}. Low values
are expected when the gas has a low density or a low kinetic temperature.
In Table 4 we give the cloud parameters derived with the first method using a uniform intrinsic 
line ratio $R$ equal to 0.5. 
We don't give the cloud size when the line ratio is larger than the maximum possible value
(observable for a point source at the beam center) or smaller than the 
intrinsic line ratio (observable for an extended source).
For $R=0.5$ estimated sizes and masses come out 
large enough that CO all disk surveys of M33 should have detected most of them. 
Volume densities are generally smaller than $10^3$~cm$^{-3}$.
Since we are searching for molecular gas around sources one may wonder whether 
the assumed value of the intrinsic line ratio is reasonable.
The intrinsic intensity ratio can in fact take values significantly larger than unity 
when the emitting regions
are warm ($T>40$~K), dense ($n_{H_2}>10^3$~cm$^{-3}$) and moderately opaque ($\tau<5$). 
However, larger values  of $R$ would imply even more massive clouds and lower densities.
Thus we conclude that it is unlikely that the intrinsic ratio $R$
is much larger than 0.5 except for the two clouds whose observed line ratio is larger than 2. 
Similarly, for s20, the intrinsic ratio is expected to be smaller than 0.5. 

Molecular masses according to the integrated CO line brightness can be computed as follows:

\begin{equation}
{M_{CO}\over M_\odot} = 41\ {X_{CO}\over 2\times 10^{20}}\ D_{arcsec}^2 {I\over 1-exp(-D_{arcsec}^2/(1.2\Theta)^2} 
\end{equation}

\noindent
where the line intensity $I$ refers to either the CO J=1-0 or J=2-1 line. 
Cloud masses derived for 
a constant conversion factor $X_{CO}=2.8\times 10^{20}$cm$^{-2}$~K$^{-1}$ are generally smaller
than virial masses (except for one source in s15). Our estimated cluster masses (see Table 4) are 
too small to account for the difference, even taking into account the incompleteness of the IMF.
In the last column we give $X_{CO}$, normalized to 10$^{20}$~cm$^{-2}$~K$^{-1}$, as the value  
needed to have the CO-luminosity mass equal to virial mass (we give the average 
between the values relative to the two CO lines): 

\begin{equation}
X_{CO}={M_{vir}\over M_{CO}} = 10 {W^2 \over D} {1-exp(-D_{arcsec}^2/(1.2\Theta_{arcsec})^2) \over I},
\end{equation}

\noindent
where the values of $I,\Theta, W$ refer to the CO line which has been used. The large cloud sizes
derived with this method for $R=0.5$ imply large cloud masses and very small star 
formation efficiencies. Gas densities are lower than expected for typical molecular cloud 
conditions (see Section 3.4 for a comparison with Milky Way clouds). 

We do not find a correlation between the cluster mass and the virial mass or between
the ratio M$_{vir}^{2-1}$/M$_{vir}^{1-0}$ and the galactocentric radius ($r_P=-0.15$). 
For this model the CO-to-H$_2$ conversion factor, $X_{CO}$, does not correlate with 
galactocentric radius ($r_P=0.34$) but there is a correlation between $X_{CO}$ and the 
cloud mass  ($r_P=0.92$), shown in Figure~2. This last correlation
implies that $X_{CO}$ increases for more massive clouds. Can this be justified? 
The $X_{CO}$ value is expected to increase as the extinction through the cloud 
or the mean volume density decreases \citep{1988ApJ...325..389M,2010arXiv1003.1340G}.
Gas densities for this model decrease as clouds get more massive, which justify the 
observed trend.   
Finally we find a marginal correlation between the cloud size and the CO linewidth, 
$D\propto W^2$, as it is observed in resolved molecular clouds  and discussed in 
Section 3.4 ($r_P$=0.61, slope
1.9$\pm$0.6 for the J=1-0 line and $r_P$=0.57 slope 1.9$\pm$0.7 for the J=2-1 line).

This method is heavily dependent on the assumption that clouds are located at the beam center,
that coincides with the peak of the IR-24~\mi\ emission. Even though the molecular gas will likely 
follow the dust distribution, the peak of the mid-IR emission depends on the intensity of
radiation field which is heating the grains. So the molecular cloud might be offset with respect to
the IR emission.

\begin{figure}
\includegraphics[width=\columnwidth]{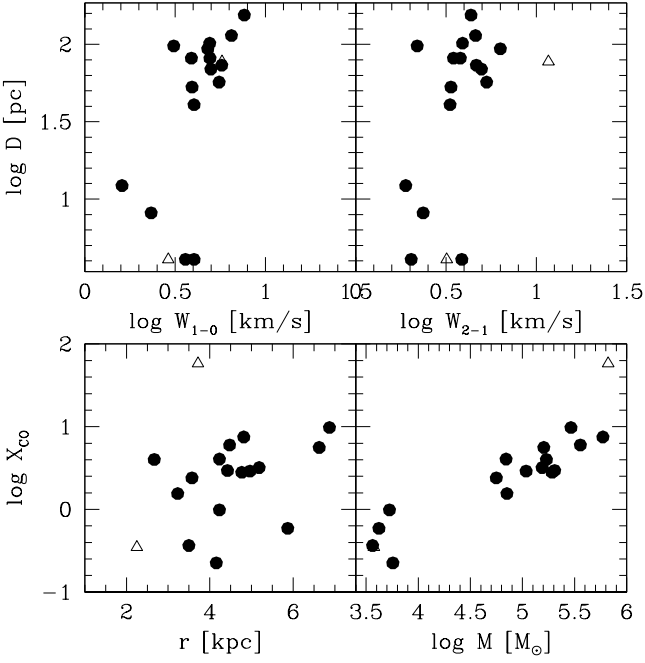}
\caption{Correlations relative to parameters derived with the first method: in the upper panels
we plot the line-width of the J=1-0 and 2-1 line versus the cloud size. 
The $X_{CO}$ factor is plotted as a function of 
galactocentric radius and of cloud mass in lower left and right panels 
respectively.}
\label{fig:f2}
\end{figure}

\subsection {Hydrostatic equilibrium}

If cloud turbulent pressure equals the ISM ambient pressure given by 
hydrostatic equilibrium we can estimate the gas volume density of the cloud. 
This is in reality a lower limit since 
cloud cores can be denser and self gravitating, and the outer parts close to equilibrium with the
surrounding ISM. From the minimum gas density we infer $D^{max}$,
the maximum cloud size, using the virial equation. As we shall remark in Section 3.4, 
it might be that only part of the cloud is
gravitationally bound and in virial equilibrium. In this case the virial equation
applies only for the higher density core with size $D<D^{max}$.

In Table 5 we give the resulting cloud parameters and below some details of this model.
We compute the ambient pressure as

\begin{equation}
P_{ISM}={\pi \over 2} G \Sigma_{g} (\Sigma_{g}+{c_g \over c_s}\Sigma_{s}) \ \ 
{\hbox{cm}}^{-3}~{\hbox{K}}
\end{equation}

\noindent
\citep{2003MNRAS.342..199C} where $\Sigma_{g,s}$ are the gas and stellar surface mass 
densities and $c_{g,s}$  are the velocity dispersion 
relative to gas and stellar disk. The velocity dispersion of the gaseous disk is  
$c_g=8$~km~s$^{-1}$. We shall use the local values of the HI column densities
for the gas surface densities (WSRT data) and a scale height of the stellar disk $z_0=0.5$~kpc 
for the stellar 
dispersion. The stellar dispersion and surface density derived through the dynamical analysis of
the rotation curve \citep{2003MNRAS.342..199C} reads:

\begin{equation}
c_s=(\pi G z_0 \Sigma_s)^{0.5} \qquad \Sigma_s=430\ exp({-R(kpc)/1.42})~M_\odot/pc^2
\end{equation}

The ISM pressure in M33 does not vary much radially since the gas surface density has a
very large scale length. However it can vary locally due to the presence of filaments
and flocculent spiral arms. For each source the turbulent pressure is 

\begin{equation}
P_{tur}=1.16 n_{H_2} T_{Turb}= 25\ n_{H_2}\ W^2(km/s) \ \ cm^{-3}~K
\end{equation}

\noindent
\citep[1.16 accounts for Helium, see also][p.37]{1978ppim.book.....S}.
The gas volume density is then:

\begin{equation}
n_{H_2}^{min}={P_{ISM}(cm^{-3}~K)\over 25\ W^2(km/s)}
\end{equation}

We shall use the smallest linewidth measured in the 1-0 or 2-1 line,
which gives the highest volume density.
We compute the maximum cloud size by
equating the virial mass to the mass in pressure equilibrium which reads:

\begin{equation}
D^{max}_{arcsec}=\sqrt{174 W^2(km/s)\over n_{H_2}^{min}}= 66 {W^2(km/s) \over \sqrt{P_{ISM}}}
\end{equation}

This relation naturally explains the size-linewidth relation $W\propto \sqrt{D}$. In fact we find
that the 2-1 linewidth (which is mostly used in this relation) correlates with the source
size, with the right slope (see Figure~3, $r_P=0.94$ slope=2.0$\pm 0.2$). 
A similar relation exists for the 1-0, but with a larger scatter ($r_P=0.81$).
The size-linewidth relation implies another important feature often mentioned about molecular
clouds, namely their rather constant column densities. For this model in fact the gas column
density reads:

\begin{equation}
{N^{min}_{H2} (cm^{-2})\over 10^{20}}= 0.126 D^{max}_{arcsec} n_{H_2}^{min} (cm^{-3}) \propto \sqrt{P_{ISM}}
\end{equation}

Values of the local ISM pressure in M33  do not vary much and this implies
almost constant column densities, as can be seen in Table 5.
We can also infer the maximum value of $X_{CO}$ from the virial mass to the CO-luminosity
mass ratio using the maximum size D:

\begin{equation}
X^{max}_{CO}={M_v\over M_{CO}} = 10 {W^2 \over D^{max}} {1-exp(-(D_{arcsec}^{max})^2/(1.2\Theta)^2 \over I}
\end{equation}

Where the values of $I,\Theta, W$ refer to the line which has been used (smaller linewidth).
The correlation between cloud mass and the CO-to-H$_2$ conversion factor is shown in
Figure 3 ($r_P=0.95$).
From the maximum size we can infer the maximum intrinsic line ratio, $R^{max}$, given 
in Table 5 for isolated sources, at the beam center.

\begin{table*}
\caption{ Clouds in pressure equilibrium with the surrounding ISM.
The CO rotational line used to infer the cloud parameters is given in column 2, the maximum value of
the intrinsic CO line ratio in column 3 and the maximum allowed
cloud diameter in column 4. The maximum cloud mass and the corresponding minimum volume
and column density of H$_2$ are given in
column 5,6 and 7 respectively. The maximum value of the CO-to-H$_2$ conversion factor is shown 
in the last column.}
\label{par4}
\begin{minipage}{\textwidth}
\begin{center}
\begin{tabular}{lccccccc}
\hline \hline
     ID  & line & R$^{max}$ & D$^{max}$ & M$^{max}$ & n$_{H_2}^{min}$ & N$^{min}_{H_2}$& X$_{CO}^{max}$ \\
         &      &           &pc & 10$^4$ M$_\odot$ & cm$^{-3}$       & 10$^{20}$ & 10$^{20}$\\
\hline \hline
    s1 &1-0 & 0.66 & 41 & 3.5 & 14. & 18 & 6.4 \\
    s2 &..  & & & & &  & \\
    s3 &1-0 & 0.66 & 106 & 35. & 8.6 & 28 & 53. \\
    s4 &..  & & & & &  & \\
    s5 &2-1 & 0.57 & 94 & 14. & 4.6 & 26 & 8.9 \\
    s6 &1-0 & 0.50 & 10 & 0.2 & 80  & 24 & 0.5 \\
    s7 &2-1 & 0.58 & 81 & 20. & 11  & 28 & 8.9 \\
    s8 &2-1 & 0.50 & 41 & 4.7 & 18  & 23 & 4.6 \\
    s9 &2-1 & 0.43 & 36 & 4.1 & 26  & 29 & 1.9 \\
   s10 &1-0 & 0.88 & 49 & 6.4 & 15  & 23 & 5.4 \\
   s11 &2-1 & 0.22 & 61 & 5.9 & 15  & 28 & 5.5 \\
   s12 &2-1 & 0.63 & 90 & 25. & 11  & 30 & 8.6 \\
   s13 &2-1 & 0.39 & 57 & 6.9 & 11  & 19 & 4.0 \\
   s14 &2-1 & 0.27 & 41 & 6.3 & 25  & 31 & 2.5 \\
   s15 &2-1 & 0.24 & 13 & 0.6 & 85  & 33 & 1.0 \\
   s15 &1-0 & 0.51 & 14 & 0.8 & 76  & 33 & 2.9 \\
   s16 &.. & & & & &  & \\
   s17 &2-1 & 0.31 & 77 & 16. & 11  & 26 & 6.7 \\
   s18 &2-1 & 0.47 & 65 & 14. & 15  & 30 & 7.0 \\
   s19 &1-0 & 0.77 & 147& 34. & 3.1 & 14 & 15. \\
   s20 &2-1 & 0.13 & 23 & 0.9 & 23. & 16 & 1.5 \\

\hline \hline
\end{tabular}
\end{center}
\end{minipage}
\end{table*}

With this method cloud sizes and masses are less dependent on the assumption of the cloud
being centered with respect to the millimeter telescope beam. Unfortunately we are able only
to derive upper limits to $M$ and $D$. In Section 3.4 we use the observed size-linewidth
relation to infer the size of self-gravitating cloud cores.
We notice that there are 3 sources with $X_{CO}^{max} \le 1.5$. Those are sources with sizes smaller
than 25~pc and small masses ($< 10^4$~M$_\odot$). Since M33 has a lower metal abundance than
the Milky Way we do not expect an overabundance of CO. Hence, these sources are
probably overluminous in CO or they are not in virial equilibrium \citep{1990ApJ...348L...9M}. 
The molecular gas might be so warm  that the J=1-0 luminosity  
is no longer proportional to the H$_2$ column density. For example for a gas at 40~K the J=1-0
line intensity is more than twice as bright as a gas at 20~K \citep[e.g.][]{1994ApJ...425..641S}.
Recent high resolution observations in the M33 outer disk have discovered clouds with high CO luminosity
\citep{2010arXiv1010.2751B} in the proximity of sources hosting massive stars. 
The very small values of R which we recover for some sources might be indicative of
an offcenter molecular cloud due to the presence of massive stars. This likely
holds for s15 where we detect two components, and for s20 where the CO observations have been 
centered on the optical HII region in the absence of a 24~\mi\ counterpart.

\subsection {Radially varying $X_{CO}$}

In Table 6 we derive the cloud properties for a radially varying $X_{CO}$ factor.
The radial scaling used takes into account the weak $X_{CO}$ metallicity dependence estimated
by \citet{2008ApJ...686..948B}:

\begin{equation}
{\hbox{log}} X_{CO}=a-0.35\times {\hbox{log}}{O\over H}\ \ \Longrightarrow\ \  
X_{CO}=2.8\times 10^{20+0.03r}
\end{equation}

\noindent 
where the average metallicity gradient O/H used is given by \citet{2010A&A...512A..63M}
and the galactocentric distance $r$ is in kpc.
The value of the constant $a$ is such that $X_{CO}$ is 2.8 10$^{20}$~cm$^{-2}$~K$^{-1}$ 
at the center of M33. The metallicity dependence of the $X_{CO}$ factor
has not yet been firmly established. We prefer to consider a shallow radial dependence
because in M33 the radial decrease of the average
radiation field, which dissociates the CO molecule, might balance the already
shallow radial metallicity gradient and keep $X_{CO}$ almost constant. 

We derive the source size by equating the
virial to the molecular mass given by the CO line luminosity:

\begin{equation}
{M_{vir}\over M_{CO}} = 406 {W^2 \over 41 X_{CO}\ I} {1-exp(-D_{arcsec}^2/(1.2\Theta)^2 \over D_{arcsec}} = 1 
\end{equation}

\noindent
From the size we infer the volume density. We then use the source size and the observed 
CO line ratio to infer the intrinsic line ratio. We take the CO line with the largest linewidth 
because it 
gives the highest gas density (see Eq.(6)). \footnote{We call some caution for this choice due
to the possibility of detecting more than one cloud in the source proximity, especially
since it is often the J=1-0 line, observed with the largest beam, which has the largest linewidth.}

For isolated clouds,
centered in the beam, the intrinsic line ratio $R$ expected for that size is also 
shown in Figure~3 and decreases moving radially outwards.

\begin{table*}
\caption{Cloud parameters for a radially varying $X_{CO}$. The CO rotational line used to 
infer the cloud parameters is given in column 2, the intrinsic CO line ratio in column 3 and the
cloud diameter in column 4. The cloud mass, the H$_2$ volume and column density are shown in 
column 5, 6 and 7 respectively.}
\label{par6}
\begin{minipage}{\textwidth}
\begin{center}
\begin{tabular}{lcccccc}
\hline \hline
     ID  & line & R & D  & M &  n$_{H_2}^{1-0}$ & N$_{H_2}$ \\
         &      &   & pc & 10$^4$ M$_\odot$ & cm$^{-3}$ & 10$^{20}$~cm$^{-2}$ \\
\hline \hline

s1 & 2-1 & 0.566&  17&  1.7&    98& 51   \\
s2 &     &      &     &     &     &      \\
s3 & 2-1 & 0.289&  1.2&  1.6&270000& 9990\\
s4 &     &      &     &     &     &      \\
s5 & 1-0 & 0.287&  21&  5.0&   160& 104  \\
s6 & 2-1 & 0.670&  56&  2.0&     3& 5.2  \\
s7 & 1-0 & 0.344&  26&  6.3&   112& 90   \\
s8 & 1-0 & 0.441&  33&  3.5&    98& 100  \\
s9 & 1-0 & 0.501&  53&  8.1&    16& 26   \\
s10& 2-1 & 0.749&  29&  4.2&    53& 47   \\
s11& 1-0 & 0.154&  13&  7.3&  1058& 424  \\
s12& 1-0 & 0.360&  33&  9.1&   100& 102  \\
s13& 1-0 & 0.340&  43&  6.4&    24& 32   \\
s14& 1-0 & 0.264&  41&  9.9&    42& 53   \\
s15& 1-0 & 0.254&  26&  2.5&    43& 35   \\
s15& 2-1 & 0.515&  19&  1.0&    46& 27   \\
s16&     &      &     &     &     &      \\
s17& 1-0 & 0.191&  19&  7.7&   360& 211  \\
s18& 1-0 & 0.329&  21&  6.9&   213& 138 \\
s19& 2-1 & 0.249&  13&  5.1&   713& 286  \\
s20& 1-0 & 0.125&  16&  2.6&   185& 91  \\

\hline \hline
\end{tabular}
\end{center}
\end{minipage}
\end{table*}

Limits to the validity of this method come from the fact that the weak metallicity dependence 
used here for $X_{CO}$ is based on CO interferometric measurements which might underestimate cloud
sizes and hence invalidate the scaling law used \citep{2001A&A...371..433I}. Metallicity 
in M33 however, has a scatter at each radius larger than the overall observed radial gradient:
a stronger metallicity dependence can be considered only if the metallicity at each
source location is known. 
This method has been based on the assumption that clouds are in virial equilibrium. In our
galaxy data suggest that clouds with masses below 10$^4$~M$_\odot$ are strongly influenced by 
nongravitational forces and virial equilibrium overestimates their masses 
\citep{1990ApJ...348L...9M}. Using only the CO observations presented in this
paper it is hard to establish if the virial equilibrium assumption is correct for clouds in
our sample. If not,  clouds might be pressure confined (as considered in Section~3.2) or related to
short-lived event (e.g. interstellar shocks).

\begin{figure}
\includegraphics[width=\columnwidth]{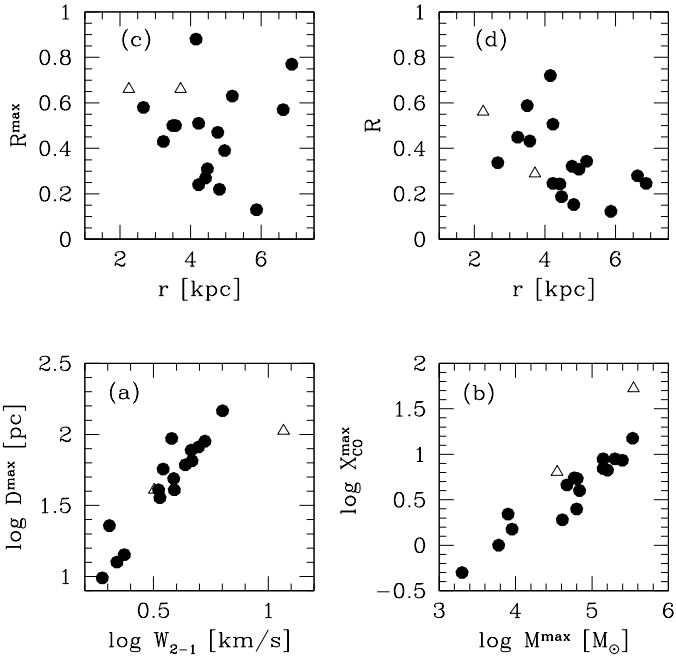}
\caption{Correlations relative to parameters derived with the  second method (a,b,c)
and with the third method (d)}
\label{fig:f3}
\end{figure}

\subsection{Comparison with spatially resolved molecular clouds}

Recent papers \citep[e.g.][]{2008ApJ...686..948B,2010ApJ...723..492R} 
summarize relevant physical properties of
Milky Way and extragalactic molecular clouds. Masses of clouds inside the solar radius
mostly range between 10$^2$ and $5\times 10^5$~M$_\odot$ and their sizes are below 25~pc.
The median value of the
average H$_2$ volume density distribution \citep[see Table 1 of ][ ]{2010ApJ...723..492R} 
is 230~cm$^{-3}$ with half of the clouds having densities between 100 and 400~cm$^{-3}$ 
and with low mass clouds having the highest densities.
The BIMA all-disk survey of M33 has identified GMCs with a completeness limit of 
10$^5$~M$_\odot$ \citep{2003ApJS..149..343E,2007ApJ...661..830R}. Hence, we expect our 
sources to be associated to smaller mass clouds,
below the completeness limit of the BIMA survey. M33 is a low-luminosity galaxy with a
lower gas and stellar surface density than the Milky Way and flocculent spiral
arms. The mass spectrum of molecular clouds in M33
is steeper than that of the Milky Way \citep{2005ASSL..327..287B} and the average cloud 
properties are more similar to those of the clouds in the outer
Galaxy. Molecular clouds in fact respond to large scale environmental variations.
In the outer Galaxy molecular clouds have smaller masses than in the inner
regions and it is still a debated question whether clouds of small mass are pressure 
bounded or in virial
equilibrium \citep{2001ApJ...551..852H,2005ASSL..327..287B,2009ApJ...699.1092H}. 

\begin{figure}
\includegraphics[width=\columnwidth]{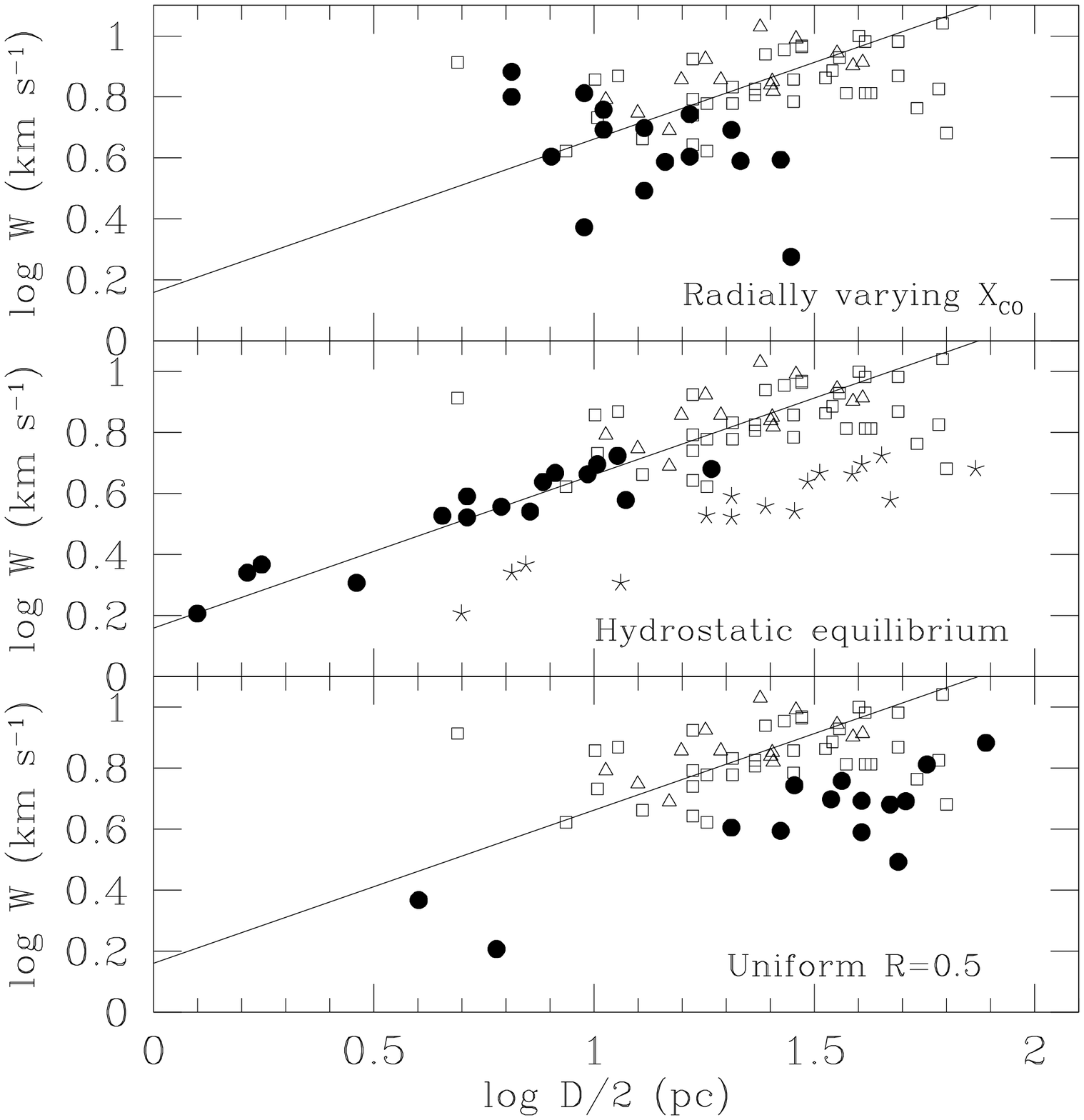}
\caption{The solid line is the fit to the size-linewidth relation for the Milky Way and 
M33 spatially resolved molecular clouds. The open symbols are for GMCs observed in M33 by
\citet{2003ApJ...599..258R} (open squares) and by \citet{1990ApJ...363..435W} (open
triangles). The filled circles are molecular clouds associated to our HII regions.
Sizes have been computed according to the uniform intrinsic line ratio in the bottom panel
and to the radially varying X$_{CO}$ model in the top panel. In the middle panel asterisks
indicate the maximum cloud size for hydrostatic equilibrium models and
filled circles the core sizes, which are 0.25 the maximum size. This scaling minimizes
the scatter around the size-linewidth relation. The linewidths used are the CO J=1-0
linewidth in the bottom panel and are given in Table 5 and Table 6 for the other two panels.
}
\label{fig:size}
\end{figure}

For a quantitative comparison between cloud properties in our sample, inferred using the three
models presented in this Section, and resolved molecular clouds, 
we shall focus on the observed size-linewidth relation:

\begin{equation}
W=1.03 \times D^{0.5}
\end{equation}

\noindent 
where $W$ is the FWHM of the CO line in km~s$^{-1}$ and D is the cloud 
diameter in parsec.
The exponent of the relation was originally derived by \citet{1987ApJ...319..730S},
while the coefficient 1.03 comes from the best fit to resolved clouds in
the Milky Way and in M33 \citep{2003ApJ...599..258R,2005ASSL..327..287B}. We use 
30$\%$ larger cloud radii for M33 data to overcome the difference in the size measurement 
technique with the Milky Way data pointed out by \citet{2003ApJ...599..258R}. 
The above relation fits the data over 
two orders of magnitude in radius, from 1 to 100~pc, and it is shown by the continuous line 
in the three panels of Figure~\ref{fig:size}. In the same Figure we plot
the data for M33 GMCs (open symbols) \citep{1990ApJ...363..435W,2003ApJ...599..258R}
and for clouds relative to HII regions in our survey (filled circles).
Cloud sizes derived using a uniform intrinsic 
line ratio R=0.5 (bottom panel) are too large to satisfy the observed size-linewidth relation.
Furthermore, cloud masses are too large to be compatible with previous surveys and
the median volume density is lower by more than a factor 10 than the value in the Milky Way.  
Considering the outer envelope of the clouds in hydrostatic
equilibrium with the ISM, we estimate higher densities. However we cannot
derive the parameters of the cloud using only this assumption 
but their limiting values. To overcome this limitation, we assume that the 
size-linewidth relation
observed in the Milky Way and in the GMCs of M33 holds for gravitationally bound 
cloud cores, and that there is an average  scaling factor of the core 
size with  respect to the maximum cloud size D$^{max}$
(given in Table 5 and plotted with asterisk symbols in the middle panel). By minimizing the
dispersion around the plotted size-linewidth relation, we derive core sizes which are
a factor 4 smaller than D$^{max}$. These are shown in the middle 
panel of Figure~\ref{fig:size} by filled circles. 
In the top panel of the figure we plot cloud sizes relative to the
radially varying X$_{CO}$ model (Table 6). We can see that despite the large scatter 
the average cloud size is smaller than that of GMCs and
not much different than what the size-linewidth relation predicts.

\section{IRAC colors and the search for embedded sources}

In the previous Section we have seen that we were able to detect CO in most star 
forming regions and
in the proximity of some evolved stars, even though CO lines are generally
weaker in the latter case. We shall now examine whether
newly born clusters can be distinguished from evolved stars using IRAC 
color-color diagrams. 
Previous studies on closer objects (like the LMC and SMC galaxies)
used IRAC color-color diagrams  to separate Young Stellar Objects (hereafter YSOs)
from stars or from evolved AGBs, and AGBs from HII regions 
\citep{2007ApJ...669..327S,2007MNRAS.374..979C,2008AIPC.1001..331M,2009AJ....138.1597B,
2009ApJS..184..172G}. Models indicate that star formation regions at different stages 
of the protostar 
to star evolutionary sequence occupy different areas  of the color-color diagrams, even
though there is overlapping and not a well defined separation between the 
various areas \citep{2006ApJS..167..256R}. Similarly \citet{2009ApJS..184..172G} have shown that
there is no simple diagnostic in color-color or color-magnitude diagrams that
can be used to uniquely separate YSOs from AGB stars. Evolved stars of different
types (Carbon-rich versus Oxygen-rich for example) lie in separate regions of the
IRAC color-color diagram but HII regions are found to lie everywhere 
\citep{2009AJ....138.1597B}. For AGB stars the [3.6]-[8.0] and [3.5]-[4.5] colors are used 
to separate
carbon stars from stars with silicate envelopes due to the presence of silicate 
and $H_2 O$ features in the spectra. These feature affect the [3.6]-[8.0] color 
excess  which seems to correlate 
with the mass-loss rate of  individual stars  and the optical thickness of 
their envelope \citep{2008AIPC.1001..331M,2006A&A...448..181G}.

\begin{figure}
\includegraphics[width=\columnwidth]{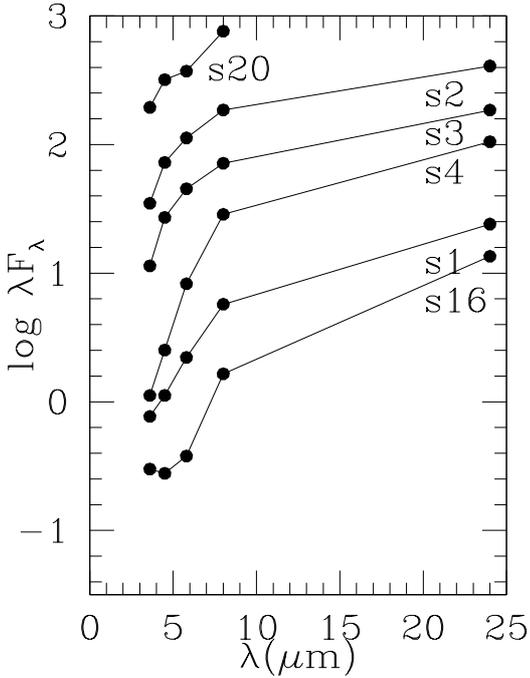}
\caption{IRAC SEDs: sources with a small flux increase between 4.5 and at 8~\mi. The plot
shows the sources from bottom to top scaling the fluxes arbitrarily so that the source at the bottom
is the one with the largest gradient between the 8 and the 24~\mi\ flux and the source at the top
is the one with the smallest gradient. Effective flux values can be recovered from Table 1.
All sources except s16 show a monotonic increase of the mid-IR flux between 3.6 and 24~\mi.
The 24~\mi\ flux for s20 is below threshold.}
\label{fig:irac1}
\end{figure}

\begin{figure}
\includegraphics[width=\columnwidth]{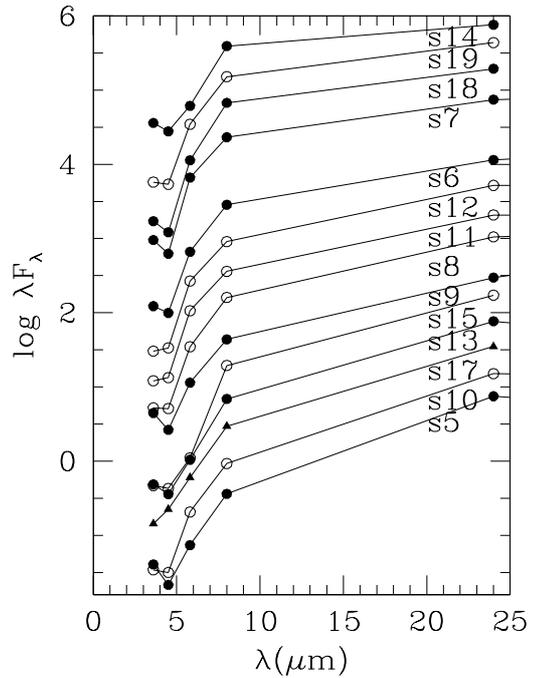}
\caption{IRAC SEDs: sources with a large flux growth between 4.5 and at 8~\mi. The plot
shows the sources from bottom to top scaling the fluxes arbitrarily so that the source at the bottom
is the one with the largest slope between the 8 and the 24~\mi\  flux, the source at the top
is the one with the smallest gradient. Effective flux values can be recovered from Table 1. Open
dots are used for sources which do not show a noticeable decrease of the flux between 3.6 and
4.5~\mi.}
\label{fig:irac2}
\end{figure}

We display the mid-IR SEDs between 3.6 and 24~\mi\ in Figures 5 and 6. 
In Figure~5
we select all sources where $\lambda F_\lambda$ varies by less than one order
of magnitude between 5.8 and 8~\mi\ , while sources with larger differences are
plotted in Figure~6. Notice that all sources in Figure \ref{fig:irac2}
have a dip at
4.5~\mi\  \citep[see also][]{2007A&A...476.1161V} except s17, which is the source  
with the lowest H$\alpha$
flux. Sources are ordered according to the SED slope between 8 and 24~\mi\ with
s5 being the source with the largest gradient and s14 the source with the smallest
one. For all sources in Figure~\ref{fig:irac2} we detect CO emission.
Sources in Figure~\ref{fig:irac1} instead have either weak or no CO emission
and no dip at 4.5~\mi. We have identified s1,s2,s3, and s20 as regions containing 
evolved variable stars. It has been shown that the MIR SED of YSOs and HII
regions have steeper slopes than evolved stars \citep{2009AJ....138.1597B,2009ApJS..184..172G},
in agreement with our finding.
This is because the SED of star forming regions in the mid-infrared peaks at wavelenghts 
longer than 24~\mi\  \citep[e.g.][ ]{2010ApJ...716..453L}. 
The CO emission associated with sources in Figure~\ref{fig:irac1}, is weak or under
our detection limit. The \citet{2007ApJ...664..850M} catalogue does not extend as far as
s4, which might be a variable associated with M33 or with our Galaxy.
The only HII region where no CO has been detected is s16, likely an evolved HII region. 
Evolved stars and evolved HII
regions show a shallow slope in the IRAC bands. The dip
at 4.5~\mi\  is present in the SEDs of HII regions 
even though some star forming
site (s17 for example), might not have it. Some YSO SEDs 
show that the dip is common but not always present \citep{2009ApJS..184..172G,2007ApJ...669..327S}.
Results from ISO spectroscopy around massive protostars have shown that CO and CO$_2$ 
absorption bands are present in the 4.5~\mi\ band \citep{2004ARA&A..42..119V}. These features
can explain the dip mostly present in the SED of star forming regions. The CO$_2$
absorption line and a time
dependent CO absorption feature can also be present  in the spectra of AGB stars 
\citep{2010ApJ...709..120M,2004A&A...417..625J} but their strengths are not sufficient
to affect notably the flux in our sample. 

\begin{figure}
\includegraphics[width=\columnwidth]{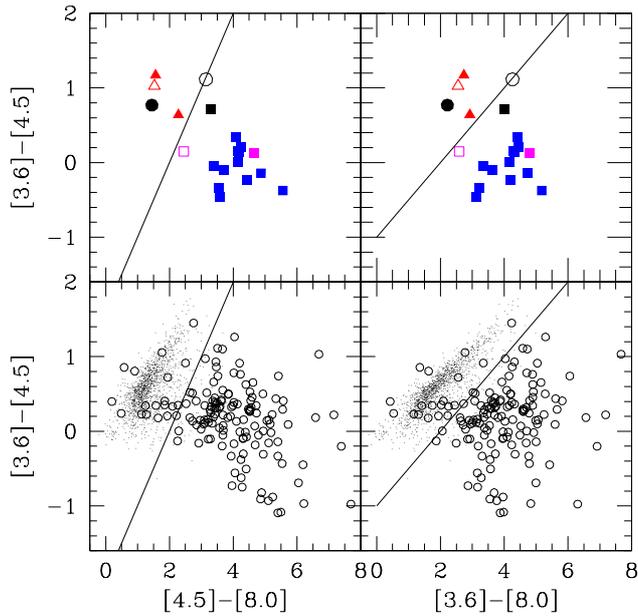}
\caption{Bottom panels display in the IRAC color-color diagrams the variable stars
detected by \citet{2007ApJ...664..850M} (small dot symbols) and young HII regions   
(the open circles; these are sources in the catalogue of \citet{2007A&A...476.1161V} 
with H$\alpha$ counterpart and not classified as variables) .
Only 3$\%$ of variable stars lie to the right of the continuous line.
In the upper panels we show the same IRAC colors but for our source sample. 
Open symbols are sources where no CO has been detected, filled symbols are sources
with CO detection. Triangles are for variable
stars. The open circle is for source 4, with uncertain classification. Squares
are for sources from 5 to 19, i.e. HII regions with 24~\mi\ counterparts. 
The filled dot is for s20, a variable star  coincident with an HII
region without 24~\mi\ emission.}
\label{fig:col1}
\end{figure}

Fig.~\ref{fig:col1} shows
all long period variables catalogued by \citet{2007ApJ...664..850M} which have been detected 
at 3.6, 4.5 and 8.0~\mi\  in the
[3.6]-[8.0]/[3.6]-[4.5] and [4.5]-[8.0]/[3.6]-[4.5] color-color diagrams.  In the same
figure we plot sources in the catalogue of \citet{2007A&A...476.1161V}
with H$\alpha$
emission, and hence are likely to be young star forming regions, and do not have
any long-term variable coincident with the extent of the 24~\mi\ emission. 
As we can see, variable stars occupy a well defined area
and HII regions have a distribution which is centered on a different region of the 
color-color diagram, with some overlap.

\begin{figure}
\includegraphics[width=\columnwidth]{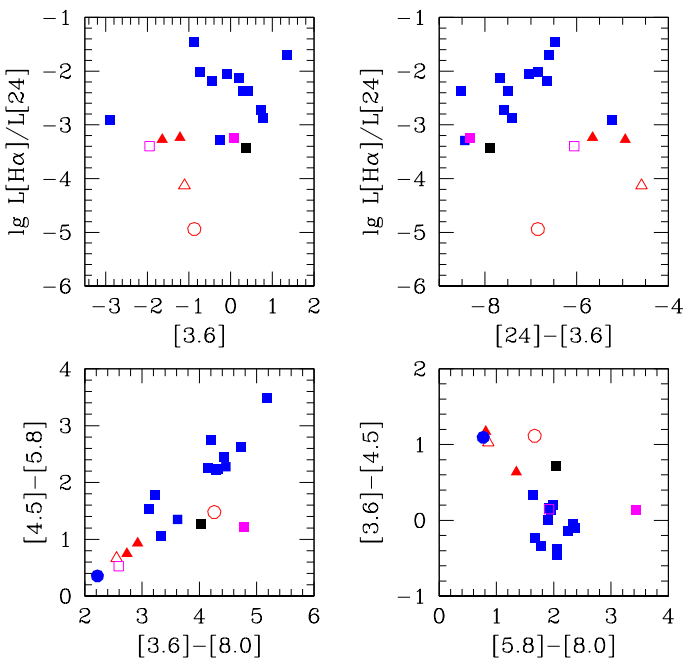}
\caption{Other colours...same symbols as in previous Figure. The source with the
highest flux at 3.6~\mi\ is s14.}
\label{fig:irac3}
\end{figure}

The two lines in Figure~\ref{fig:col1} have slopes 1 and 0.5 respectively and separate the 
region populated by evolved stars.
Only about 3$\%$ of the evolved stars lie to the right of the lines (about 10$\%$ 
of the IR selected HII regions lie to their left).
The sources considered in this paper well represent the two distributions.
Our HII regions with a detectable CO, 24~\mi\ and H$\alpha$ emission lie in
a well defined region and not in the
upper left corner where most of the evolved stars lie. Close to the dividing
line we find s17, an
embedded HII region candidate without H$\alpha$ and very low UV fluxes, 
and s16, the HII region devoid of CO.
We cannot exclude that s16, s17 host an evolved star which did not meet the
classification schemes used (actually they are both within a few parsec of
a non-point source variable according to \citet{2006MNRAS.371.1405H}).
 The probability for an AGB star to
encounter a SF region is low in the solar neighborhood  \citep{1994ApJ...421..605K}
but it can be appreciable along spiral arms and filaments.
The colours of s20, an evolved
HII region with no 24~\mi\ counterpart and associated to an evolved variable star
has IRAC colors similar to variable stars.

\begin{figure}
\includegraphics[width=\columnwidth]{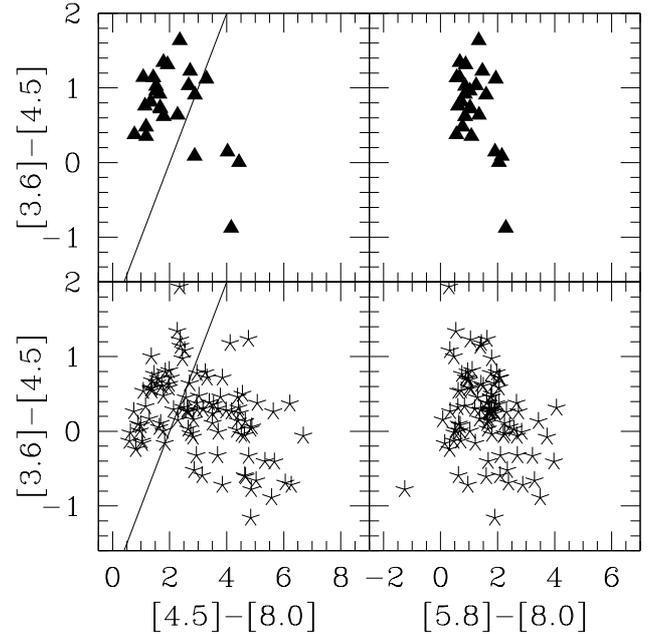}
\caption{Star symbols in the lower panels are for 24~\mi\ sources with no H$\alpha$ counterpart
which have been not identified by \citet{2007ApJ...664..850M} as variable stars. 
Filled triangles in the upper panels are sources in \citet{2007A&A...476.1161V} catalogue which 
host variable stars according to \citet{2007ApJ...664..850M}. }
\label{fig:col2}
\end{figure}

In Figure~\ref{fig:irac3} we display other variables and colors, even outside the IRAC
range. In each panel the young HII regions where dust and CO emission
have been detected occupy well defined areas. For example all of them except one (s14)
have the 3.6~\mi\ magnitude above -1 and have [24]-[3.6]$<-6.5$. Similarly their IRAC
color [5.8]-[8.0] is $>1.6$. There is also a nice correlation between
the [4.5]-[5.8] color and the [3.6]-[8.0] color with 
the young HII regions where dust and CO emission have been detected lying at 
[4.5]-[5.8]$>1$. 
Triangles in the upper panels of Figure~\ref{fig:col2} indicate the colors of 24~\mi\
sources in the catalogue of \citet{2007A&A...476.1161V} with no H$\alpha$ counterpart
which ``host'' a variable star i.e.
at least one classified variable is located within the 24~\mi\
emission boundary.   Only 25 out of 125
sources in the catalogue of  \citet{2007A&A...476.1161V} with no H$\alpha$ counterpart 
and detected in IRAC bands have been classified as  
variable stars. Most of these lie to the left of the 
lines drawn in Figure~\ref{fig:col2}. 
Variable sources which lie to the right of the line may contain
both an evolved star and an embedded HII region. In the bottom
panels we show sources with no H$\alpha$ counterpart according to the catalogue of 
\citet{2007A&A...476.1161V} not associated to known
variable stars. The number of these sources
which lie to the left of the line exceed by far the percentage expected if
these sources have similar colors to the visible HII regions. 
Either embedded stars occupy a wider area of the IRAC color-color diagram than HII
regions or there are unidentified AGBs, variables in crowded 
regions which escaped the classification by \citet{2007ApJ...664..850M}.
The majority of sources to the right of the line are good candidate for being young 
star forming regions. Only 5 of these sources are associated with known GMCs.  
The associated molecular clouds might be of smaller mass, like the ones detected 
in the CO survey presented here.

\section{The cluster birthline test}

\begin{figure}
\includegraphics[width=\columnwidth]{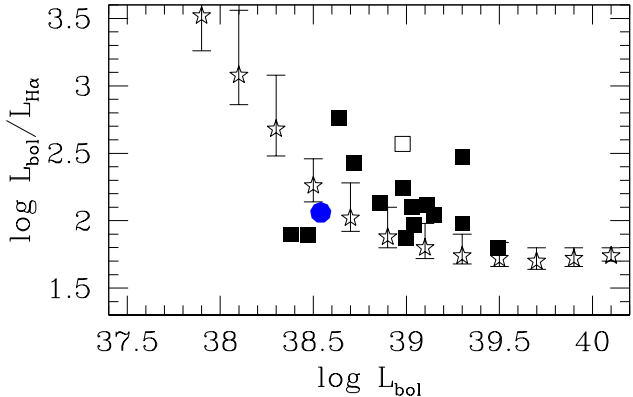}
\caption{Sources on the cluster birthline. Star symbols indicate the birthline relative to a 
stochastically sampled universal IMF with their 1-$\sigma$ dispersion.
Open symbols are for sources where no CO has been detected, filled symbols are sources
with a CO detection. The filled dot is for s20 which does not have IR emission.}
\label{fig:birth}
\end{figure}

To verify the young age of our sources we will check if they lie close to the cluster birthline,
which is the line in the L$_{bol}$--L$_{bol}$/L$_{H\alpha}$ plane around which young clusters lie
\citep{2009A&A...495..479C}. Older clusters should be located
above the birthline because of the faster fading of the H$\alpha$
luminosity compared to the bolometric luminosity. Leakage of ionizing photons or the ejection or the delayed formation of 
massive stars can also move the 
clusters above the birthline. In Figure~\ref{fig:birth} we show with star symbols 
the birthline relative to a 
stochastically sampled universal IMF with a Salpeter slope -2.35 at the high-mass end. Stochasticity
allows a spread in the values of L$_{bol}$/L$_{H\alpha}$ for a given L$_{bol}$: for a given
bolometric luminosity we can have a fully populated IMF up to a certain mass M$_{max}$, or an IMF
populated
only up to M$_*<$ M$_{max}$, plus a brighter outlier. 
The errorbars in each
bin indicate the dispersion due to bright outliers that can form for the given L$_{bol}$.
The highest simulated L$_{bol}$/L$_{H\alpha}$ value is for a fully populated IMF case. 
We see that there are two sources (s12,s16) which are clearly above the
upper 3-$\sigma$ boundary. These might be more evolved sources or leaking ionizing radiation. 
Since we have not found CO emission from s16, it is likely that it is an evolving source.
The source s12 is the brightest in the IR and has a strong CO emission: 
thus, its position can be explained if it is in the process of forming its massive stellar population or if it is leaking 
ionizing photons. The strong 8~\mi\
emission places s12 in the upper right corner of the [3.6]-[8.0]--[4.5]-[5.8] color plot.
All the other sources cannot be found below the birthline
if the IMF assumptions are correct. Our sources are all compatible with a random
sampled IMF model including the two sources, s13 and s19, close to the lower boundary which might 
host outlier stars.

\section{Conclusions}

We have investigated the nature of faint 24~\mi\ sources with a weak
or no H$\alpha$ counterpart. These could be regions with  
embedded clusters still associated to molecular gas, or
evolved stars with dusty envelopes. To distinguish between these two possibilities,
we have used published catalogues of variable stars 
in M33 and carried out
deep observations of the CO J=1-0 and 2-1 line with the IRAM 30-mt  telescope to search
for molecular gas associated with 20 sources. The main results are:

\begin{itemize}

\item
Deep pointed CO observations
have revealed the presence of CO around 17 of the 20 sources included in our sample.
The weakness of the CO lines around most of our sources
is indicative of the existence of a large population of faint CO clouds in M33. 
This is in agreement with the results of the CO survey of 
\citet{2010arXiv1003.3222G} 
and with the steeper molecular mass spectrum relative to our 
Galaxy \citep{2005ASSL..327..287B}.

\item
The estimate of cloud masses is uncertainsince our pointed observations are deeper than 
any existing CO map.  Considering clouds in virial equilibrium
and a weak radial dependence of the CO-to-H$_2$ conversion factor, we find molecular
gas masses in the range $10^4$ -- $10^5$~M$_\odot$. In this case, cloud sizes agree
on average with those predicted by the size-linewidth relation of the Milky Way and
resolved GMCs in other galaxies, but have large dispersions. We have also estimated limits to
molecular gas masses and sizes in the case where cloud outer envelopes are in pressure 
equilibrium with the surrounding medium.
This model reproduces  the observed size-linewidth relation correctly if
gravitationally bound cloud cores are a factor 4 smaller in size than the outer envelopes.

\item
The intrinsic CO J=2-1 to J=1-0 line ratio is generally
small, $\sim$0.4, if one makes the assumption that clouds are within a few
parsecs from the beam center. 
The CO lines are weaker at the location of evolved stars than around star 
forming regions, as is the H$\alpha$ and 70~\mi\ emission. Stellar clusters associated
to clouds in our sample are young since they lie along the cluster birthline.

\item
In the IRAC color-color diagrams AGB variable stars and 
HII regions occupy distinct areas even though there is some overlap: most of the AGBs
have [3.6]-[4.5] color in the range 0 to 1.5 and lie around a line of slope 0.5
in the [3.6]-[8.0]--[3.6]-[4.5] plane. Most of the 24~\mi\ sources associated to
HII regions have higher values for the [4.5]-[8.0] or [3.6]-[8.0] color than variable
stars. The IRAC colors
of our 20 selected sources are consistent with these distributions. They show a
linear correlation between the [4.5]-[5.8] color and the [3.6]-[8.0] color with
the AGBs having the lowest values. 
The IRAC colors confirm the presence of AGBs in 4 sources of our sample
whose variability was previously detected either in infrared or optical surveys.

\item
Using IRAC color-color diagrams we also predict
that in the sample of 24~\mi\ sources without H$\alpha$ counterpart there are still
unidentified evolved stars other than young clusters. The young clusters might
lack H$\alpha$ emission because they are of small mass or because they are 
very compact, and deeply embedded. Additional sensitive CO 
observations and high resolution far-infrared images will help to identify 
unambiguously the properties of the young star forming sites in M33.

\end{itemize}

\begin{acknowledgements}
We would like to thank the staff in Granada
for their assistance during the IRAM 30-mt observations and the referee for comments
to the original manuscript.
The study presented in this paper has been supported by the European Community 
Framework Programme 7, Advanced Radio Astronomy in Europe, grant agreement no.: 227290.
\end{acknowledgements}

\bibliography{edv}

\begin{thebibliography}{58}
\expandafter\ifx\csname natexlab\endcsname\relax\def\natexlab#1{#1}\fi

\bibitem[{{Bigiel} {et~al.}(2010){Bigiel}, {Bolatto}, {Leroy}, {Blitz},
  {Walter}, {Rosolowsky}, {Lopez}, \& {Plambeck}}]{2010arXiv1010.2751B}
{Bigiel}, F., {Bolatto}, A., {Leroy}, A., {et~al.} 2010, ArXiv e-prints

\bibitem[{{Blitz} \& {Rosolowsky}(2005)}]{2005ASSL..327..287B}
{Blitz}, L. \& {Rosolowsky}, E. 2005, in Astrophysics and Space Science
  Library, Vol. 327, The Initial Mass Function 50 Years Later, ed.
  E.~{Corbelli}, F.~{Palla}, \& H.~{Zinnecker}, 287--+

\bibitem[{{Bolatto} {et~al.}(2008){Bolatto}, {Leroy}, {Rosolowsky}, {Walter},
  \& {Blitz}}]{2008ApJ...686..948B}
{Bolatto}, A.~D., {Leroy}, A.~K., {Rosolowsky}, E., {Walter}, F., \& {Blitz},
  L. 2008, \apj, 686, 948

\bibitem[{{Buchanan} {et~al.}(2009){Buchanan}, {Kastner}, {Hrivnak}, \&
  {Sahai}}]{2009AJ....138.1597B}
{Buchanan}, C.~L., {Kastner}, J.~H., {Hrivnak}, B.~J., \& {Sahai}, R. 2009,
  \aj, 138, 1597

\bibitem[{{Buckalew} {et~al.}(2006){Buckalew}, {Kobulnicky}, {Darnel},
  {Polomski}, {Gehrz}, {Humphreys}, {Woodward}, {Hinz}, {Engelbracht},
  {Gordon}, {Misselt}, {P{\'e}rez-Gonz{\'a}lez}, {Rieke}, {Willner}, {Ashby},
  {Barmby}, {Pahre}, {Roellig}, {Devereux}, {Loon}, \&
  {Brandl}}]{2006ApJS..162..329B}
{Buckalew}, B.~A., {Kobulnicky}, H.~A., {Darnel}, J.~M., {et~al.} 2006, \apjs,
  162, 329

\bibitem[{{Cohen} {et~al.}(2007){Cohen}, {Green}, {Meade}, {Babler},
  {Indebetouw}, {Whitney}, {Watson}, {Wolfire}, {Wolff}, {Mathis}, \&
  {Churchwell}}]{2007MNRAS.374..979C}
{Cohen}, M., {Green}, A.~J., {Meade}, M.~R., {et~al.} 2007, \mnras, 374, 979

\bibitem[{{Corbelli}(2003)}]{2003MNRAS.342..199C}
{Corbelli}, E. 2003, \mnras, 342, 199

\bibitem[{{Corbelli} {et~al.}(2010){Corbelli}, {Giovanardi}, \&
  {Grossi}}]{2010arXiv1011.1097C}
{Corbelli}, E., {Giovanardi}, C., \& {Grossi}, M. 2010, ArXiv e-prints

\bibitem[{{Corbelli} {et~al.}(2009){Corbelli}, {Verley}, {Elmegreen}, \&
  {Giovanardi}}]{2009A&A...495..479C}
{Corbelli}, E., {Verley}, S., {Elmegreen}, B.~G., \& {Giovanardi}, C. 2009,
  \aap, 495, 479

\bibitem[{{Crosthwaite} \& {Turner}(2007)}]{2007AJ....134.1827C}
{Crosthwaite}, L.~P. \& {Turner}, J.~L. 2007, \aj, 134, 1827

\bibitem[{{Engargiola} {et~al.}(2003){Engargiola}, {Plambeck}, {Rosolowsky}, \&
  {Blitz}}]{2003ApJS..149..343E}
{Engargiola}, G., {Plambeck}, R.~L., {Rosolowsky}, E., \& {Blitz}, L. 2003,
  \apjs, 149, 343

\bibitem[{{Fazio} {et~al.}(2004){Fazio}, {Hora}, {Allen}, {Ashby}, {Barmby},
  {Deutsch}, {Huang}, {Kleiner}, {Marengo}, {Megeath}, {Melnick}, {Pahre},
  {Patten}, {Polizotti}, {Smith}, {Taylor}, {Wang}, {Willner}, {Hoffmann},
  {Pipher}, {Forrest}, {McMurty}, {McCreight}, {McKelvey}, {McMurray}, {Koch},
  {Moseley}, {Arendt}, {Mentzell}, {Marx}, {Losch}, {Mayman}, {Eichhorn},
  {Krebs}, {Jhabvala}, {Gezari}, {Fixsen}, {Flores}, {Shakoorzadeh}, {Jungo},
  {Hakun}, {Workman}, {Karpati}, {Kichak}, {Whitley}, {Mann}, {Tollestrup},
  {Eisenhardt}, {Stern}, {Gorjian}, {Bhattacharya}, {Carey}, {Nelson},
  {Glaccum}, {Lacy}, {Lowrance}, {Laine}, {Reach}, {Stauffer}, {Surace},
  {Wilson}, {Wright}, {Hoffman}, {Domingo}, \& {Cohen}}]{2004ApJS..154...10F}
{Fazio}, G.~G., {Hora}, J.~L., {Allen}, L.~E., {et~al.} 2004, \apjs, 154, 10

\bibitem[{{Freedman} {et~al.}(1991){Freedman}, {Wilson}, \&
  {Madore}}]{1991ApJ...372..455F}
{Freedman}, W.~L., {Wilson}, C.~D., \& {Madore}, B.~F. 1991, \apj, 372, 455

\bibitem[{{Gardan} {et~al.}(2007){Gardan}, {Braine}, {Schuster}, {Brouillet},
  \& {Sievers}}]{2007A&A...473...91G}
{Gardan}, E., {Braine}, J., {Schuster}, K.~F., {Brouillet}, N., \& {Sievers},
  A. 2007, \aap, 473, 91

\bibitem[{{Gil de Paz} {et~al.}(2007){Gil de Paz}, {Boissier}, {Madore},
  {Seibert}, {Joe}, {Boselli}, {Wyder}, {Thilker}, {Bianchi}, {Rey}, {Rich},
  {Barlow}, {Conrow}, {Forster}, {Friedman}, {Martin}, {Morrissey}, {Neff},
  {Schiminovich}, {Small}, {Donas}, {Heckman}, {Lee}, {Milliard}, {Szalay}, \&
  {Yi}}]{2007ApJS..173..185G}
{Gil de Paz}, A., {Boissier}, S., {Madore}, B.~F., {et~al.} 2007, \apjs, 173,
  185

\bibitem[{{Glover} \& {Mac Low}(2010)}]{2010arXiv1003.1340G}
{Glover}, S.~C.~O. \& {Mac Low}, M. 2010, ArXiv e-prints

\bibitem[{{Gratier} {et~al.}(2010){Gratier}, {Braine}, {Rodriguez-Fernandez},
  {Schuster}, {Kramer}, {Xilouris}, {Tabatabaei}, {Henkel}, {Corbelli},
  {Israel}, {van der Werf}, {Calzetti}, {Garcia-Burillo}, {Sievers}, {Combes},
  {Wiklind}, {Brouillet}, {Herpin}, {Bontemps}, {Aalto}, {Koribalski}, {van der
  Tak}, {Wiedner}, {Roellig}, \& {Mookerjea}}]{2010arXiv1003.3222G}
{Gratier}, P., {Braine}, J., {Rodriguez-Fernandez}, N.~J., {et~al.} 2010, ArXiv
  e-prints

\bibitem[{{Greenawalt}(1998)}]{1998PhDT........16G}
{Greenawalt}, B.~E. 1998, PhD thesis, AA(NEW MEXICO STATE UNIVERSITY)

\bibitem[{{Groenewegen}(2006)}]{2006A&A...448..181G}
{Groenewegen}, M.~A.~T. 2006, \aap, 448, 181

\bibitem[{{Groenewegen} {et~al.}(2007){Groenewegen}, {Wood}, {Sloan},
  {Blommaert}, {Cioni}, {Feast}, {Hony}, {Matsuura}, {Menzies}, {Olivier},
  {Vanhollebeke}, {van Loon}, {Whitelock}, {Zijlstra}, {Habing}, \&
  {Lagadec}}]{2007MNRAS.376..313G}
{Groenewegen}, M.~A.~T., {Wood}, P.~R., {Sloan}, G.~C., {et~al.} 2007, \mnras,
  376, 313

\bibitem[{{Grossi} {et~al.}(2010){Grossi}, {Corbelli}, {Giovanardi}, \&
  {Magrini}}]{2010arXiv1006.1281G}
{Grossi}, M., {Corbelli}, E., {Giovanardi}, C., \& {Magrini}, L. 2010, ArXiv
  e-prints

\bibitem[{{Gruendl} \& {Chu}(2009)}]{2009ApJS..184..172G}
{Gruendl}, R.~A. \& {Chu}, Y. 2009, \apjs, 184, 172

\bibitem[{{Hartman} {et~al.}(2006){Hartman}, {Bersier}, {Stanek}, {Beaulieu},
  {Kaluzny}, {Marquette}, {Stetson}, \&
  {Schwarzenberg-Czerny}}]{2006MNRAS.371.1405H}
{Hartman}, J.~D., {Bersier}, D., {Stanek}, K.~Z., {et~al.} 2006, \mnras, 371,
  1405

\bibitem[{{Heyer} {et~al.}(2009){Heyer}, {Krawczyk}, {Duval}, \&
  {Jackson}}]{2009ApJ...699.1092H}
{Heyer}, M., {Krawczyk}, C., {Duval}, J., \& {Jackson}, J.~M. 2009, \apj, 699,
  1092

\bibitem[{{Heyer} {et~al.}(2001){Heyer}, {Carpenter}, \&
  {Snell}}]{2001ApJ...551..852H}
{Heyer}, M.~H., {Carpenter}, J.~M., \& {Snell}, R.~L. 2001, \apj, 551, 852

\bibitem[{{Heyer} {et~al.}(2004){Heyer}, {Corbelli}, {Schneider}, \&
  {Young}}]{2004ApJ...602..723H}
{Heyer}, M.~H., {Corbelli}, E., {Schneider}, S.~E., \& {Young}, J.~S. 2004,
  \apj, 602, 723

\bibitem[{{Hoopes} \& {Walterbos}(2000)}]{2000ApJ...541..597H}
{Hoopes}, C.~G. \& {Walterbos}, R.~A.~M. 2000, \apj, 541, 597

\bibitem[{{Israel} \& {Baas}(2001)}]{2001A&A...371..433I}
{Israel}, F.~P. \& {Baas}, F. 2001, \aap, 371, 433

\bibitem[{{Justtanont} {et~al.}(2004){Justtanont}, {de Jong}, {Tielens},
  {Feuchtgruber}, \& {Waters}}]{2004A&A...417..625J}
{Justtanont}, K., {de Jong}, T., {Tielens}, A.~G.~G.~M., {Feuchtgruber}, H., \&
  {Waters}, L.~B.~F.~M. 2004, \aap, 417, 625

\bibitem[{{Kastner} \& {Myers}(1994)}]{1994ApJ...421..605K}
{Kastner}, J.~H. \& {Myers}, P.~C. 1994, \apj, 421, 605

\bibitem[{{Lawton} {et~al.}(2010){Lawton}, {Gordon}, {Babler}, {Block},
  {Bolatto}, {Bracker}, {Carlson}, {Engelbracht}, {Hora}, {Indebetouw},
  {Madden}, {Meade}, {Meixner}, {Misselt}, {Oey}, {Oliveira}, {Robitaille},
  {Sewilo}, {Shiao}, {Vijh}, \& {Whitney}}]{2010ApJ...716..453L}
{Lawton}, B., {Gordon}, K.~D., {Babler}, B., {et~al.} 2010, \apj, 716, 453

\bibitem[{{Leitherer} {et~al.}(1999){Leitherer}, {Schaerer}, {Goldader},
  {Gonz{\'a}lez Delgado}, {Robert}, {Kune}, {de Mello}, {Devost}, \&
  {Heckman}}]{1999ApJS..123....3L}
{Leitherer}, C., {Schaerer}, D., {Goldader}, J.~D., {et~al.} 1999, \apjs, 123,
  3

\bibitem[{{Magrini} {et~al.}(2010){Magrini}, {Stanghellini}, {Corbelli},
  {Galli}, \& {Villaver}}]{2010A&A...512A..63M}
{Magrini}, L., {Stanghellini}, L., {Corbelli}, E., {Galli}, D., \& {Villaver},
  E. 2010, \aap, 512, A63+

\bibitem[{{Makovoz} \& {Khan}(2005)}]{2005ASPC..347...81M}
{Makovoz}, D. \& {Khan}, I. 2005, in Astronomical Society of the Pacific
  Conference Series, Vol. 347, Astronomical Data Analysis Software and Systems
  XIV, ed. {P.~Shopbell, M.~Britton, \& R.~Ebert}, 81--+

\bibitem[{{Maloney}(1990)}]{1990ApJ...348L...9M}
{Maloney}, P. 1990, \apjl, 348, L9

\bibitem[{{Maloney} \& {Black}(1988)}]{1988ApJ...325..389M}
{Maloney}, P. \& {Black}, J.~H. 1988, \apj, 325, 389

\bibitem[{{Marengo} {et~al.}(2010){Marengo}, {Evans}, {Barmby}, {Bono},
  {Welch}, \& {Romaniello}}]{2010ApJ...709..120M}
{Marengo}, M., {Evans}, N.~R., {Barmby}, P., {et~al.} 2010, \apj, 709, 120

\bibitem[{{Marengo} {et~al.}(2008){Marengo}, {Reiter}, \&
  {Fazio}}]{2008AIPC.1001..331M}
{Marengo}, M., {Reiter}, M., \& {Fazio}, G.~G. 2008, in American Institute of
  Physics Conference Series, Vol. 1001, Evolution and Nucleosynthesis in AGB
  Stars, ed. {R.~Guandalini, S.~Palmerini, \& M.~Busso}, 331--338

\bibitem[{{Martin} {et~al.}(2005){Martin}, {Fanson}, {Schiminovich},
  {Morrissey}, {Friedman}, {Barlow}, {Conrow}, {Grange}, {Jelinsky},
  {Milliard}, {Siegmund}, {Bianchi}, {Byun}, {Donas}, {Forster}, {Heckman},
  {Lee}, {Madore}, {Malina}, {Neff}, {Rich}, {Small}, {Surber}, {Szalay},
  {Welsh}, \& {Wyder}}]{2005ApJ...619L...1M}
{Martin}, D.~C., {Fanson}, J., {Schiminovich}, D., {et~al.} 2005, \apjl, 619,
  L1

\bibitem[{{McQuinn} {et~al.}(2007){McQuinn}, {Woodward}, {Willner}, {Polomski},
  {Gehrz}, {Humphreys}, {van Loon}, {Ashby}, {Eicher}, \&
  {Fazio}}]{2007ApJ...664..850M}
{McQuinn}, K.~B.~W., {Woodward}, C.~E., {Willner}, S.~P., {et~al.} 2007, \apj,
  664, 850

\bibitem[{{Oka} {et~al.}(1998){Oka}, {Hasegawa}, {Hayashi}, {Handa}, \&
  {Sakamoto}}]{1998ApJ...493..730O}
{Oka}, T., {Hasegawa}, T., {Hayashi}, M., {Handa}, T., \& {Sakamoto}, S. 1998,
  \apj, 493, 730

\bibitem[{{Rieke} {et~al.}(2004){Rieke}, {Young}, {Engelbracht}, {Kelly},
  {Low}, {Haller}, {Beeman}, {Gordon}, {Stansberry}, {Misselt}, {Cadien},
  {Morrison}, {Rivlis}, {Latter}, {Noriega-Crespo}, {Padgett}, {Stapelfeldt},
  {Hines}, {Egami}, {Muzerolle}, {Alonso-Herrero}, {Blaylock}, {Dole}, {Hinz},
  {Le Floc'h}, {Papovich}, {P{\'e}rez-Gonz{\'a}lez}, {Smith}, {Su}, {Bennett},
  {Frayer}, {Henderson}, {Lu}, {Masci}, {Pesenson}, {Rebull}, {Rho}, {Keene},
  {Stolovy}, {Wachter}, {Wheaton}, {Werner}, \&
  {Richards}}]{2004ApJS..154...25R}
{Rieke}, G.~H., {Young}, E.~T., {Engelbracht}, C.~W., {et~al.} 2004, \apjs,
  154, 25

\bibitem[{{Robitaille} {et~al.}(2006){Robitaille}, {Whitney}, {Indebetouw},
  {Wood}, \& {Denzmore}}]{2006ApJS..167..256R}
{Robitaille}, T.~P., {Whitney}, B.~A., {Indebetouw}, R., {Wood}, K., \&
  {Denzmore}, P. 2006, \apjs, 167, 256

\bibitem[{{Roman-Duval} {et~al.}(2010){Roman-Duval}, {Jackson}, {Heyer},
  {Rathborne}, \& {Simon}}]{2010ApJ...723..492R}
{Roman-Duval}, J., {Jackson}, J.~M., {Heyer}, M., {Rathborne}, J., \& {Simon},
  R. 2010, \apj, 723, 492

\bibitem[{{Rosolowsky} {et~al.}(2003){Rosolowsky}, {Engargiola}, {Plambeck}, \&
  {Blitz}}]{2003ApJ...599..258R}
{Rosolowsky}, E., {Engargiola}, G., {Plambeck}, R., \& {Blitz}, L. 2003, \apj,
  599, 258

\bibitem[{{Rosolowsky} {et~al.}(2007){Rosolowsky}, {Keto}, {Matsushita}, \&
  {Willner}}]{2007ApJ...661..830R}
{Rosolowsky}, E., {Keto}, E., {Matsushita}, S., \& {Willner}, S.~P. 2007, \apj,
  661, 830

\bibitem[{{Sakamoto} {et~al.}(1994){Sakamoto}, {Hayashi}, {Hasegawa}, {Handa},
  \& {Oka}}]{1994ApJ...425..641S}
{Sakamoto}, S., {Hayashi}, M., {Hasegawa}, T., {Handa}, T., \& {Oka}, T. 1994,
  \apj, 425, 641

\bibitem[{{Simon} {et~al.}(2007){Simon}, {Bolatto}, {Whitney}, {Robitaille},
  {Shah}, {Makovoz}, {Stanimirovi{\'c}}, {Barb{\'a}}, \&
  {Rubio}}]{2007ApJ...669..327S}
{Simon}, J.~D., {Bolatto}, A.~D., {Whitney}, B.~A., {et~al.} 2007, \apj, 669,
  327

\bibitem[{{Solomon} {et~al.}(1987){Solomon}, {Rivolo}, {Barrett}, \&
  {Yahil}}]{1987ApJ...319..730S}
{Solomon}, P.~M., {Rivolo}, A.~R., {Barrett}, J., \& {Yahil}, A. 1987, \apj,
  319, 730

\bibitem[{{Sorai} {et~al.}(2001){Sorai}, {Hasegawa}, {Booth}, {Rubio},
  {Morino}, {Bronfman}, {Handa}, {Hayashi}, {Nyman}, {Oka}, {Sakamoto}, {Seta},
  \& {Usuda}}]{2001ApJ...551..794S}
{Sorai}, K., {Hasegawa}, T., {Booth}, R.~S., {et~al.} 2001, \apj, 551, 794

\bibitem[{{Spitzer}(1978)}]{1978ppim.book.....S}
{Spitzer}, L. 1978, {Physical processes in the interstellar medium}, ed.
  {Spitzer, L.}

\bibitem[{{Thilker} {et~al.}(2005){Thilker}, {Hoopes}, {Bianchi}, {Boissier},
  {Rich}, {Seibert}, {Friedman}, {Rey}, {Buat}, {Barlow}, {Byun}, {Donas},
  {Forster}, {Heckman}, {Jelinsky}, {Lee}, {Madore}, {Malina}, {Martin},
  {Milliard}, {Morrissey}, {Neff}, {Schiminovich}, {Siegmund}, {Small},
  {Szalay}, {Welsh}, \& {Wyder}}]{2005ApJ...619L..67T}
{Thilker}, D.~A., {Hoopes}, C.~G., {Bianchi}, L., {et~al.} 2005, \apjl, 619,
  L67

\bibitem[{{van Dishoeck}(2004)}]{2004ARA&A..42..119V}
{van Dishoeck}, E.~F. 2004, \araa, 42, 119

\bibitem[{{Verley} {et~al.}(2009){Verley}, {Corbelli}, {Giovanardi}, \&
  {Hunt}}]{2009A&A...493..453V}
{Verley}, S., {Corbelli}, E., {Giovanardi}, C., \& {Hunt}, L.~K. 2009, \aap,
  493, 453

\bibitem[{{Verley} {et~al.}(2010){Verley}, {Corbelli}, {Giovanardi}, \&
  {Hunt}}]{2010A&A...510A..64V}
{Verley}, S., {Corbelli}, E., {Giovanardi}, C., \& {Hunt}, L.~K. 2010, \aap,
  510, A64+

\bibitem[{{Verley} {et~al.}(2007){Verley}, {Hunt}, {Corbelli}, \&
  {Giovanardi}}]{2007A&A...476.1161V}
{Verley}, S., {Hunt}, L.~K., {Corbelli}, E., \& {Giovanardi}, C. 2007, \aap,
  476, 1161

\bibitem[{{Werner} {et~al.}(2004){Werner}, {Roellig}, {Low}, {Rieke}, {Rieke},
  {Hoffmann}, {Young}, {Houck}, {Brandl}, {Fazio}, {Hora}, {Gehrz}, {Helou},
  {Soifer}, {Stauffer}, {Keene}, {Eisenhardt}, {Gallagher}, {Gautier}, {Irace},
  {Lawrence}, {Simmons}, {Van Cleve}, {Jura}, {Wright}, \&
  {Cruikshank}}]{2004ApJS..154....1W}
{Werner}, M.~W., {Roellig}, T.~L., {Low}, F.~J., {et~al.} 2004, \apjs, 154, 1

\bibitem[{{Wilson} \& {Scoville}(1990)}]{1990ApJ...363..435W}
{Wilson}, C.~D. \& {Scoville}, N. 1990, \apj, 363, 435

\end{thebibliography}

\end{document}